\documentclass[a4paper,twocolumn,aps,prd,nolongbibliography,superscriptaddress,showpacs,showkeys,amsmath,amssymb,floatfix,nofootinbib]{revtex4-1}
\usepackage{graphicx}
\usepackage{lmodern}

\usepackage[T1]{fontenc}
\usepackage[utf8]{inputenc}
\usepackage{amsmath}
\usepackage{color}
\usepackage[unicode=true,pdfusetitle,
 bookmarks=true,bookmarksnumbered=true,bookmarksopen=true,bookmarksopenlevel=1,
 breaklinks=false,pdfborder={0 0 0},backref=false,colorlinks=true]
 {hyperref}
\hypersetup{
 citecolor=blue,filecolor=blue,linkcolor=blue,urlcolor=blue}
\usepackage[normalem]{ulem}
\makeatletter

\pdfpageheight\paperheight
\pdfpagewidth\paperwidth

 \@ifundefined{textcolor}{}
 {%
   \definecolor{BLACK}{gray}{0}
   \definecolor{WHITE}{gray}{1}
   \definecolor{RED}{rgb}{1,0,0}
   \definecolor{green}{rgb}{0,0.7,0}
   \definecolor{BLUE}{rgb}{0,0,1}
   \definecolor{CYAN}{cmyk}{1,0,0,0}
   \definecolor{MAGENTA}{cmyk}{0,1,0,0}
   \definecolor{YELLOW}{cmyk}{0,0,1,0}
   \definecolor{SIENA}{rgb}{0.772,0.541,0.243}
 }

\newcommand{\be}{\begin{equation}}
\newcommand{\ee}{\end{equation}}
\newcommand{\ba}{\begin{eqnarray}}
\newcommand{\ea}{\end{eqnarray}}

\newcommand{\bk}{\boldsymbol{k}}
\newcommand{\bx}{\boldsymbol{x}}

\newcommand{\bhn}{\boldsymbol{\hat{n}}}
\newcommand{\ve}{\varepsilon}
\newcommand{\dd}{\partial}

\renewcommand{\[}{\begin{equation}}
\renewcommand{\]}{\end{equation}}
\newcommand{\cvis}{c_\text{vis}}
\newcommand{\cs}{c_\text{s}}
\newcommand{\cvisz}{c_\text{vis0}}
\newcommand{\csz}{c_\text{s0}}

\newcommand{\camb}{{\sc camb}}
\newcommand{\cosmomc}{{\sc CosmoMC}}

\usepackage{array}
\makeatother

\begin{document}

\title{Constraints on dark-matter properties from large-scale structure}\thanks{Based in part on observations obtained with Planck (http://www.esa.int/Planck), an ESA science mission with instruments and contributions directly funded by ESA Member States, NASA, and Canada.}

\author{Martin Kunz}
\affiliation{D\'epartment de Physique Th\'eorique and Center for Astroparticle Physics,
Universit\'e de Gen\`eve, Quai E. Ansermet 24, CH-1211 Gen\`eve 4, Switzerland}

\author{Savvas Nesseris}
\affiliation{Instituto de F\'isica Te\'orica UAM-CSIC, Universidad Auton\'oma de Madrid,
Cantoblanco, 28049 Madrid, Spain}

\author{Ignacy Sawicki}
\affiliation{D\'epartment de Physique Th\'eorique and Center for Astroparticle Physics,
Universit\'e de Gen\`eve, Quai E. Ansermet 24, CH-1211 Gen\`eve 4, Switzerland}

\begin{abstract}
We use large-scale cosmological observations to place constraints on the dark-matter pressure, sound speed and viscosity, and infer a limit on the mass of warm-dark-matter particles. Measurements of the cosmic microwave background  anisotropies constrain the equation of state and sound speed of the dark matter at last scattering at the per mille level. Since the redshifting of collisionless particles universally implies that these quantities scale like $a^{-2}$ absent shell crossing, we infer that today  $w_{\rm (DM)}< 10^{-10.0}$, $c_{\rm s,(DM)}^2 < 10^{-10.7}$ and $c_{\rm vis, (DM)}^{2} < 10^{-10.3}$ at the $99\%$ confidence level. This very general bound can be translated to model-dependent constraints on dark-matter models: for warm dark matter these constraints imply $m> 70$ eV, assuming it decoupled while relativistic around the same time as the neutrinos; for a cold relic, we show that $m>100$ eV. We separately constrain the properties of the DM fluid on linear scales at late times, and find upper bounds  $c_{\rm s, (DM)}^2<10^{-5.9}$, $c_{\rm vis, (DM)}^{2} < 10^{-5.7}$, with no detection of non-dust properties for the DM.
\end{abstract}
\maketitle

% ----------------------------------------------------------------------------------------------------------

\section{Introduction}

Dark matter (DM) is one of the key ingredients in the current standard model of cosmology $\Lambda$CDM, and is thought to make up about 26\% of the energy density today \cite{Ade:2015xua}. It is necessary for the formation of structure by gravitational clustering and is needed to explain the rotation curves of galaxies and the motion of galaxies in clusters. In the concordance cosmological model, $\Lambda$CDM, dark matter is modeled as dust --- pressureless matter moving on geodesics. A typical concrete realization of this kind of dark matter is provided by weakly interacting massive particles (WIMPs) with masses of the order of 100 GeV.

However, many years of direct and indirect searches have been unable to provide a clear detection of any particles that make up the dark matter. An important goal is therefore to place as many constraints as possible on the different quantities that characterize its physical nature. For example, the Bullet cluster places limits on the self interaction cross section of dark matter particles to $\sigma/m < 1 \,\mathrm{cm}^2 \mathrm{g}^{-1}$ \cite{Markevitch:2003at}. If dark matter particles are fermionic and  too light, $m \lesssim 400$ eV then their Fermi pressure does not allow structure to form (the Tremaine-Gunn bound \cite{Tremaine:1979we}). Other constraints come from the clustering seen in the Lyman-$\alpha$ forest, for which a comparison with hydrodynamical simulations leads to a bound of $m>3.3$ keV at $2\sigma$ \cite{Viel:2013apy}. There are however also claims from X-ray observations concerning the detection of a 3.55 keV line that might be due to the two-body decay of a dark matter particle with a mass of 7.1 keV (see for example \cite{Iakubovskyi:2015wma} for a recent review). For more details and further bounds see e.g.\ \cite{Bertone:2004pz,Boyarsky:2012rt,Adhikari:2016bei}.

In this paper we study the constraints that can be placed on the fluid aspect of the dark matter, i.e. its pressure, sound speed and viscosity, from cosmological observations on large scales, and the implications of these general results for a broad class of particle dark matter models. Large-scale observations in cosmology have the advantage of requiring only linear physics, which makes them an especially clean and highly successful probe \cite{Adam:2015rua}. As we will see, the bounds from observations of the anisotropies in the cosmic microwave background (CMB), the lensing of the CMB and the weak lensing of galaxies are comparable to those obtained from physics on smaller scales. The limits we obtain are highly model independent and robust, and they come from high redshifts (close to last scattering) as well as low redshifts (lensing of the CMB). Where comparable, our results agree with constraints obtained recently in ref.~\cite{Thomas:2016iav} and previously with older datasets in \cite{Calabrese:2009zza,Xu:2013mqe}.

The paper is organized as follows: We start by describing the way we model the dark matter and how this is connected to the dark matter mass. We also discuss how we implement this in the Boltzmann code \camb\ \cite{Lewis:1999bs} and how we set the initial conditions. In section \ref{sec:results} we briefly review the different data sets used, before presenting the results. We then discuss the implications for dark matter physics and conclude.

% ----------------------------------------------------------------------------------------------------------

\section{Describing Dark Matter}
\label{sec:description}
\subsection{Evolution}

In this paper, we will assume that, at all times relevant for observations, the dark matter is decoupled from the visible sector (baryons, photons and neutrinos) in any manner except for gravity. This allows us to restrain its evolution and its effect on observables to that which is allowed by the conservation of the energy-momentum tensor (EMT) for dark matter. This assumption means that we do not consider, for example, models where dark matter is either metastable or continues to annihilate to radiation at a sufficient rate to affect the temperature of the plasma. For constraints on such effects see e.g.\ ref.~\cite{Yang:2015cva}.

This means that all kinds of dark-matter that we will cover can be described by the standard conservation equations for a general matter source as given by ref.~\cite{Ma:1995ey}, the notational conventions of which we adopt here. In particular, on the level of the cosmological background, the DM energy density $\rho$ evolves according to
\begin{equation}
\dot\rho + 3H(1+w)\rho = 0,
\end{equation}
where the overdot signifies differentiation w.r.t.\ conformal time $\tau$ and $H\equiv \dot{a}/a$ is the conformal Hubble parameter. The equation of state $w$ will in this paper denote the equations of state of the DM, rather than any dark energy. We assume that we can consistently neglect vector and tensor perturbations, so that we can consider only the scalar modes. Thus, on the level of linear perturbations, a conserved EMT must satisfy \cite{Ma:1995ey}:
\begin{align}
	\dot{\delta} &+ (1+w)\left( \theta + \frac{\dot{h}}{2} \right) + 3H\left(\frac{\delta p}{\delta\rho} -w \right)\delta =0 \,,\\
	\dot{\theta} &+H(1-3w)\theta + \frac{\dot{w}}{1+w} \theta - \frac{\delta p/\delta\rho}{1+w} k^2 \delta +k^2\sigma = 0 \,,\notag
\end{align}
where we have presented the equations in synchronous gauge and in a frame comoving with a pressureless dust component, i.e.\ the choice of variables made in the \camb\ numerical code \cite{Lewis:1999bs} that we use to obtain the results in this paper.%
\footnote{When $1+w\approx 0$, there are some technical issues related to the observer choice for a general EMT, and it is not always possible to choose the comoving frame consistently \cite{Kunz:2006wc,Nesseris:2006er,Sawicki:2012pz}. This will not be an issue here.}%

Given our freeze-out/energy-conservation assumption, the model is specified by supplying a DM equation of state, $w$, and relations associating the pressure perturbation $\delta p$ and scalar anisotropic stress $\sigma$ to the variables being evolved dynamically, $\delta, \theta$ or the gravitational potentials.

Frequently these relations are taken from perfect-fluid hydrodynamics, as in the case of CDM. However, one cannot necessarily assume that the dark matter is an ideal fluid with a natural suppression of higher-order terms in a gradient expansion. The DM particles interact very rarely compared to the timescale of cosmological evolution and thus cannot establish thermodynamical equilibrium which would lead to such a hierarchy, but rather free-stream. Instead, the above relations are obtained by solving the Boltzmann equation for the particle distribution (for more details see Appendix~\ref{sec:Boltzmann}), typically through a multipole moment decomposition. Then one finds that each higher moment is suppressed with respect to the lower one by the ratio of the particle kinetic energy to its mass. This means that hydrodynamics is a terrible approximation when the DM is relativistic (just as in the case of neutrinos) and the full set of coupled moment equations must be solved, but the moment expansion can be truncated when the DM is non-relativistic.

Since the dark matter does need to be non-relativistic at least at the present time to allow for the formation of galaxies, we will employ a truncation of the multipole expansion which was introduced in ref.~\cite{Hu:1998kj,Hu:1998tj}, the so-called $\cvis$ parametrization.  This parametrization relates the pressure perturbation to the dynamically evolved variables through the rest-frame sound-speed $\cs$
\begin{equation}
	\delta p = \cs^2 \delta\rho - \dot\rho (\cs^2-c_\text{a}^2)\theta/k^2,
\end{equation}
where the adiabatic sound speed is $c_\text{a}^2 \equiv (w\rho)\dot{}/\dot\rho$. In addition, the anisotropic stress $\sigma$ is assumed to evolve through the phenomenological equation
\begin{equation}
	\dot{\sigma}+3H\frac{c_\text{a}^2}{w}\sigma = \frac{4}{3}\frac{\cvis^2}{1+w}(2\theta+\dot{h}+6\dot{\eta}) \label{eq:sigmaEvol}\, ,
\end{equation}
where $\cvis^2$ is a new viscosity parameter. As discussed in ref.~\cite{Oldengott:2014qra}, such a parametrization in the limit $\cvis^2=0$ only restores the hydrodynamical limit of the Boltzmann hierarchy when the multipoles higher than the quadrupole are unpopulated as an initial condition, which is not the case for a real relativistic species. On the other hand, in the relativistic limit $w=\cs^2=\cvis^2=1/3$, this set of equations is also missing the input from the higher multipoles and therefore is not a very realistic representation. However, since dark matter must be non-relativistic today we expect that the effect of the higher multipoles is sufficiently suppressed so as not to make a significant correction to observables. We thus treat $\cvis^2$ as a proxy for the size of the higher multipoles. If we were to find that the data support $\cvis^2\gg\cs^2$, then a more precise investigation of the higher moments is necessary.

All that remains therefore is to specify the time evolution of three parameters: $w, \cs^2, \cvis^2$. We will study two parameterizations:
\begin{enumerate}
	\item \emph{Initially relativistic DM}. We implement time-varying $w, \cs^2, \cvis^2$ interpolating between relativistic and non-relativistic behavior. This is a physically motivated parametrization, based on the redshifting of momenta of collisionless particles and it allows us to obtain very general constraints on warm-dark-matter-type (WDM) scenarios.
		
	\item \emph{Constant parametrization}: We take all parameters $w, \cs^2, \cvis^2$ to be constant. This will allow us to ascertain the maximum values that these parameters are allowed to take and also infer the behavior of the DM fluid at late times. Comparing the two parameterizations will reveal from which redshift and therefore due to which physics the constraints arise. This kind of constraints were recently obtained also by \cite{Thomas:2016iav}.
	
\end{enumerate}

\begin{figure*}[!t]
\centering
\includegraphics[width=1\columnwidth]{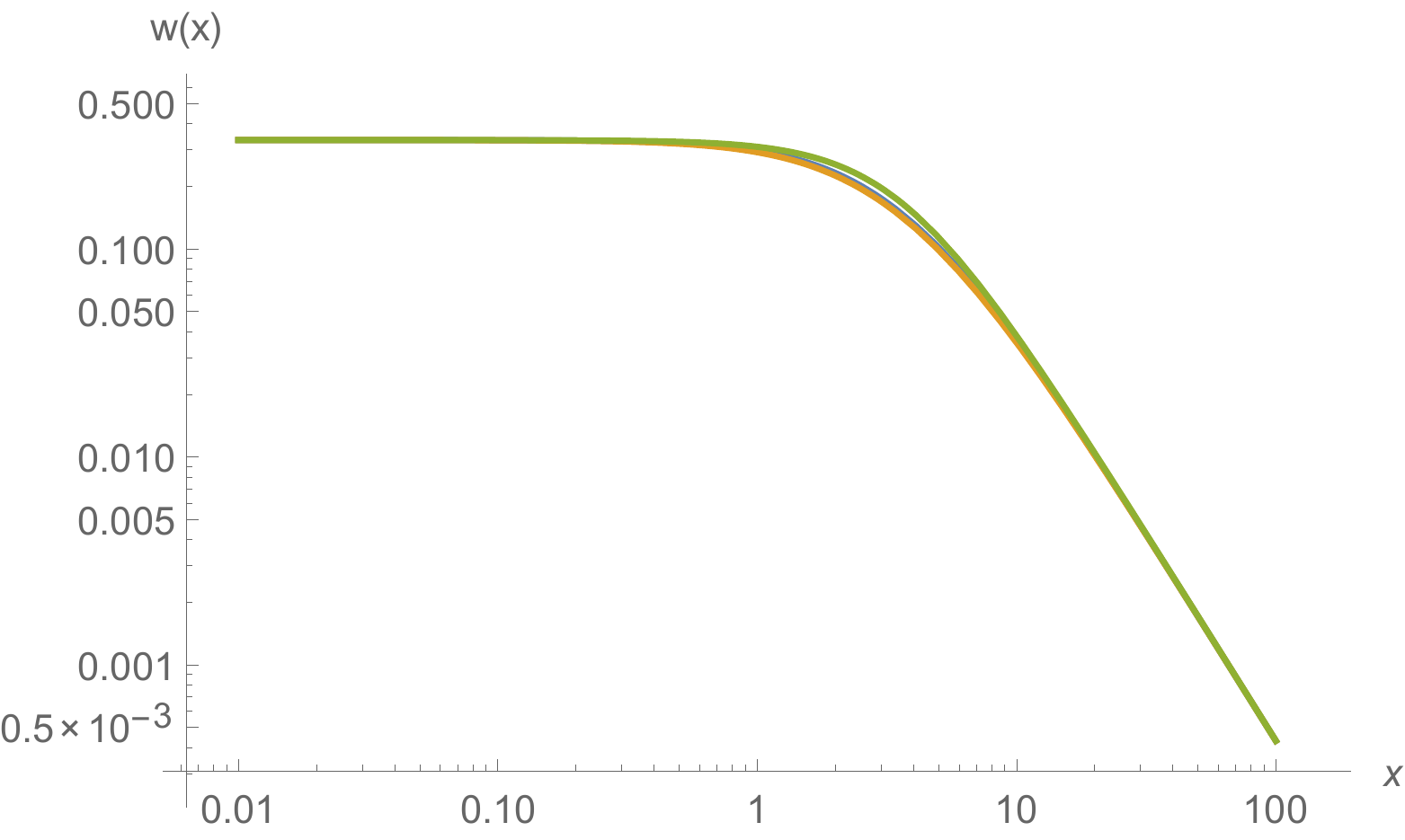}
\includegraphics[width=1\columnwidth]{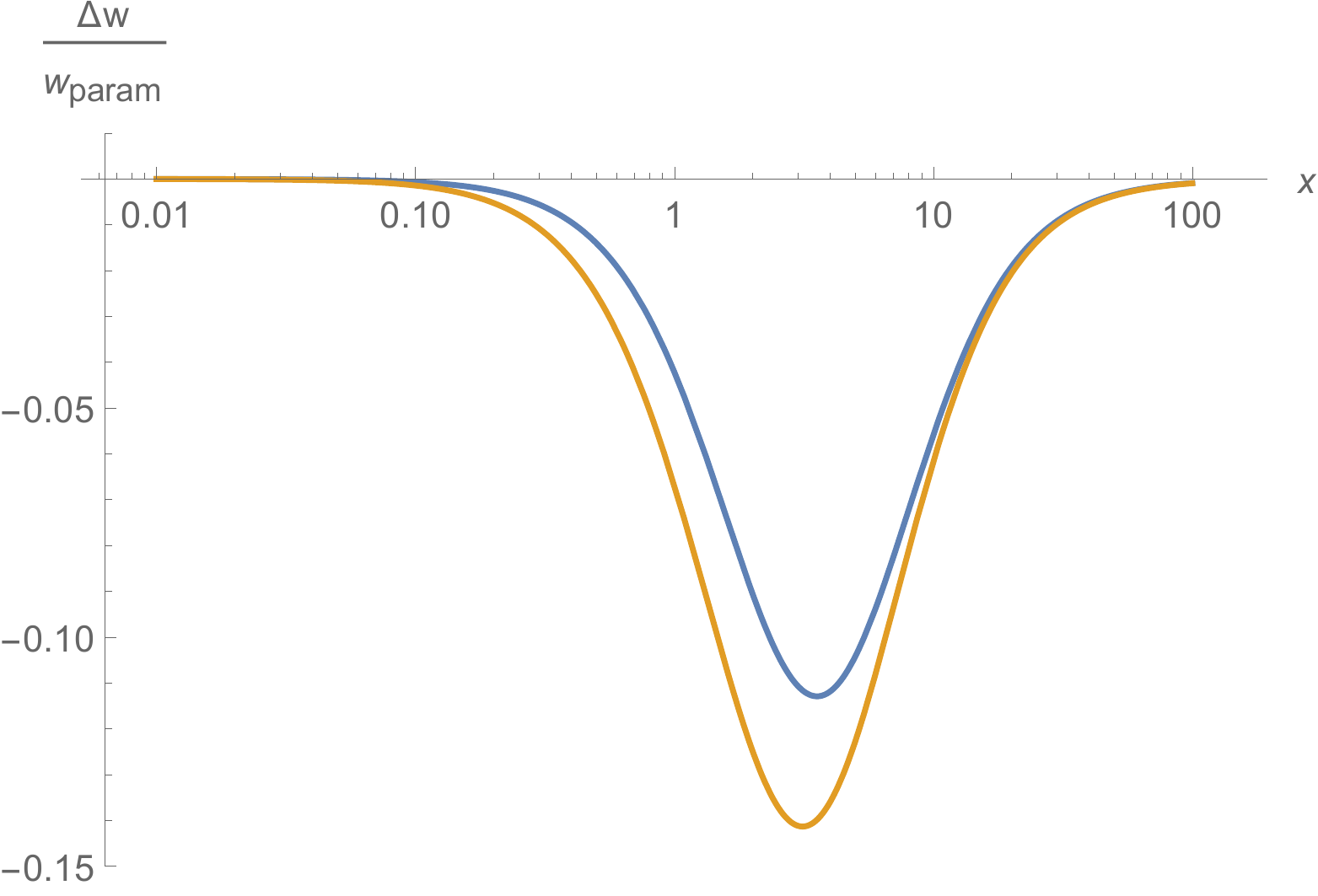}
\caption{\label{fig:eos} Left panel: The effective equation of state parameter $w$ for fermionic particles as a function of a rescaled scale factor $x$ (orange) and the parameterization \eqref{eq:paramer-w} (green).\\
Right panel: The relative difference in $w$ for fermions (blue) and bosons (orange) and the parameterization \eqref{eq:paramer-w}. The asymptotic behavior matches, while there is a 10-15\% difference around $x=3$, when the species are transitioning between relativistic and non-relativistic. For $x\gtrsim 30$ the parameterization is accurate to 1\%, and it is always conservative.}
\end{figure*}

As we review in more detail in Appendix~\ref{sec:Boltzmann}, the distribution functions for DM evolves according to the Vlasov equation. Provided it interacts sufficiently frequently, it can thermalize and be well described by a hydrodynamical perfect fluid. However, once freeze-out occurs at DM temperature $T_\text{dec}$, the full Boltzmann hierarchy must in principle be evolved since higher moments are only suppressed by the ratio of the typical kinetic energy to total particle energy. We thus need to model an initial relativistic limit, with $w=\cs^2=\frac{1}{3}$ and unsuppressed higher multipoles of the hierarchy.

On the other hand, the phase-space distribution scales in a universal manner following freeze-out, since the particles are now only redshifting with the expansion of the universe but no longer interacting. Thus, whatever the precise DM generation/freeze-out scenario, once non-relativistic and collisionless, the kinetic energy, and therefore also the pressure, redshifts as $a^{-2}$ while the higher multipoles become increasingly irrelevant. This allows us to employ a parametrization that is independent of the precise model of dark matter: we parameterize the evolution of all of $w, \cs^2$ and $\cvis^2$ using the same functional form
\begin{equation}
	F(x) = \frac{1}{3+x^2} \label{eq:paramer-w}\,,
\end{equation}
where $x\equiv a/\sqrt{\alpha}$ with $\alpha$ one of $w_0, \csz^2, \cvisz^2$, the value of these fluid parameters today. The function $F$ interpolates between $\frac{1}{3}$ at early times and $\frac{\alpha}{a^2}$ at late times. We compare how well this analytic approximation compares with the full numerical calculation for the equation of state in fig.~\ref{fig:eos}. Here, it suffices to say that, in the case of a species that was thermally distributed until decoupling while relativistic, this parametrization is conservative. The interpretation of a constraint on the fluid parameters in terms the DM particle mass depends on the freeze-out scenario and is discussed in section~\ref{sec:discussion}. We also note that $x$ at late times is approximately proportional to $m/T$, the ratio of the effective temperature of the DM to its mass.

Note that, once the DM becomes non-relativistic, the higher moments of the Boltzmann hierarchy decay away faster than $\cs^2$ and $w$. Nonetheless, since we use a phenomenological approximation to the full hierarchy through the $\cvis^2$ parameter, and are mostly looking for an upper bound on any effects from the higher moments, for $\cvis^2$ we employ the same parametrization \eqref{eq:paramer-w}. With this parametrization, a preference in the data for $\cvisz^2>\csz^2$ would imply that the higher moments are larger than the lower and our approximation cannot be employed.

The discussion above strictly speaking applies until shell crossing, at which point, the velocities of the fluid elements become multivalued and must be re-averaged. This changes the hydrodynamical parameters for the fluid (e.g.\ by introducing pressure from the velocity dispersion) and therefore would break the $a^{-2}$ scaling. This effect occurs at low redshifts at scales that become non-linear, and thus our analysis should not be sensitive to it. Nonetheless, we employ an alternative, constant parameterization to gauge the magnitude of any potential such effect in the data. We will demonstrate that it is small.

\subsection{Initial Conditions}

In addition to the evolution equations, appropriate initial conditions must be chosen for the evolution of the modes. We generalize the prescription developed in ref.~\cite{Ma:1995ey}, and extended in ref.~\cite{Ballesteros:2010ks}.

This prescription assumes that after starting from pure adiabatic inflationary initial conditions, the configuration of each mode evolves towards an attractor solution. This attractor is the appropriate initial condition valid at extremely superhorizon scales, when the species are not in causal contact and pressure support is absent.  Such an attractor can only exist when the universe is in a scaling solution (in particular, radiation domination) and the DM parameters $w,\cs^2,\cvis^2$ are constant.

We leave the details for appendix \ref{app:IC}. Carrying out a full parameter space investigation requires that correct initial conditions be set. However, it turns out that for the parameter values allowed by the data, the observables are not sensitive to the initial conditions. Thus the posterior distribution also is insensitive to the choice of ICs.

\subsection{Implementation}

We implement this extended DM model in the \camb\ numerical code by exploiting the \emph{dark degeneracy} \cite{Kunz:2007rk}. Note that an alternative implementation of non-cold dark matter is available for the CLASS Boltzmann code \cite{Lesgourgues:2011rh}. We modify the \camb\ code by combining our extended DM and the cosmological constant (``$\Lambda$'') into a single fluid and removing the CDM component in \camb, repurposing the modification we performed for ref.~\cite{Kunz:2015oqa}.  We modify the density and the equation of state of the DM to take into account the constant contribution of $\Lambda$. We thus define a density fraction of the combined generalized DM and cosmological constant
\begin{equation}
\Omega_X(a)  = \Omega_\Lambda(a) + \Omega_c(a) \, .
\end{equation}
The combined fluid then evolves with an equation of state
\begin{equation}
1+w_X(a) = (1+w)\frac{\Omega_c(a)}{\Omega_X(a)} \,.\label{eq:w0}
\end{equation}
As one should expect, when $\Lambda$ is subdominant, the equation of state is just that of the DM. This takes care of the modifications in the background.

The cosmological constant carries no perturbations and has equation of state $w_\Lambda=-1$. We can thus use the standard perturbation equations for dark energy already implemented in \camb\ to describe the combined DM/$\Lambda$ fluid, using $\Omega_X$ and $w_X$ as the dark energy density fraction and equation of state but not adjusting at all any of the parameters $\cs^2, \cvis^2$ or $c_\text{a}^2$. The only point of care is in implementing equation \eqref{eq:sigmaEvol}, where the $w$ in the friction term always is the $w$ of the DM component alone. The fact that the $\sigma$ evolution equation is not adjusted ``automatically'' is a result of its not arising from a well-defined generally covariant model.

% ----------------------------------------------------------------------------------------------------------

\section{Results\label{sec:results}}
For our analysis, we have  modified the \camb/\cosmomc\ public codes \cite{Lewis:1999bs,Lewis:2002ah} to implement the changes described in section~\ref{sec:description}. Our model contains at most three extra parameters compared to the concordance $\Lambda$CDM model, although we will fix some of them in some runs.

We perform the analysis by constraining our model using the 2015 Planck CMB likelihoods \cite{Aghanim:2015xee}, in some cases adding the likelihood for the gravitational lensing of the CMB from the trispectrum \cite{Ade:2015zua}. In order to provide a reasonable representation of the degeneracies, we also always include distance data together with each of the perturbation-related data sets. Therefore, we have included the BAO measurements from CMASS and LOWZ of Ref.~\cite{Anderson:2013zyy}, the 6DF measurement from Ref.~\cite{Beutler:2011hx}, the MGS measurement from Ref.~\cite{Ross:2014qpa} and the JLA SNe Ia catalog from \cite{Betoule:2014frx}, all readily available in the \cosmomc\ code. We do not include any measurements of the Hubble constant $H_0$, apart from a uniform prior $0.4\leq h \leq 1.0$.

In addition for some of the runs, we include the ultraconservative cut of the galaxy weak lensing shear (WL) correlation function from the CFHTLenS survey \cite{Heymans:2013fya}. As is well known, these results are mildly incompatible with Planck when $\Lambda$CDM is assumed for the cosmology \cite{Ade:2015xua}. We investigate the extent to which an extended DM model might resolve the tension between these data while noting that a recent re-analysis of CFHTLenS data using 3D cosmic shear seems to suggest that the discrepancy can be resolved by an appropriate cut of the non-linear scales and the introduction of a bias for photometric redshifts \cite{Kitching:2016hvn}. The science verification data release from the Dark Energy Survey is compatible with both the data sets \cite{Abbott:2015swa}.

\subsection{Extended DM and Halofit}

The effect of introducing a non-zero DM sound speed is to prevent clustering inside the Jeans length, thus cutting off the matter power spectrum inside this scale. If the sound speed is high enough, or increases sufficiently rapidly with redshift, fluctuations can be sufficiently erased so as not to allow non-linear structure. This would prevent collapsed objects such as galaxies from ever forming.

\begin{figure}
\centering
\vspace{0cm}\rotatebox{0}{\vspace{0cm}\hspace{0cm}\resizebox{0.49\textwidth}{!}{\includegraphics{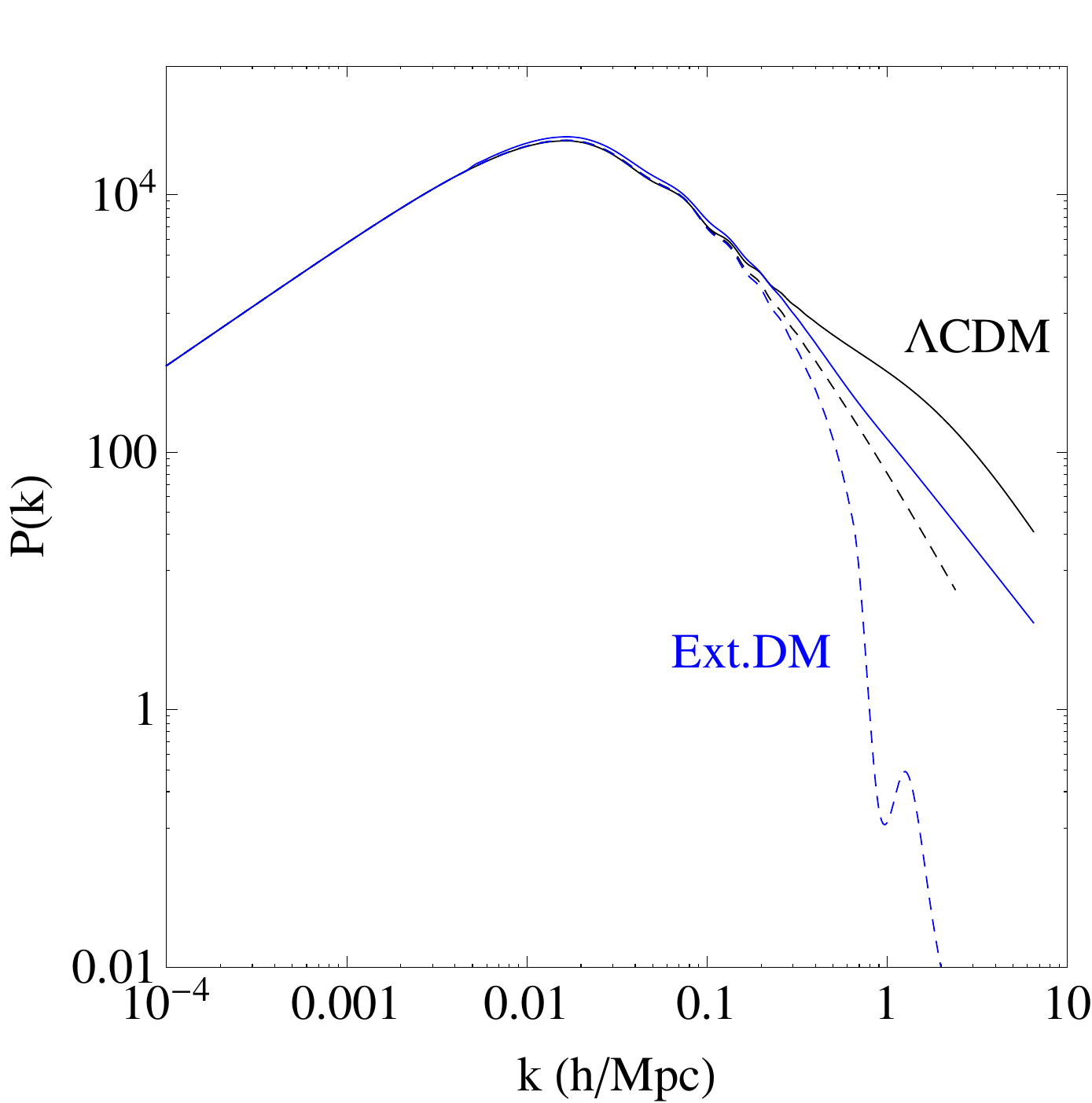}}}
\caption{The matter power spectrum $P(k)$ for the extended DM model (blue lines) versus $\Lambda$CDM (black lines) with or without halofit (solid and dashed lines respectively) for the parameters $w=c_s^2=10^{-11}$, $\cvis^2=10^{-50}$ and $\Omega_{m,0}=0.30$.\label{fig:halo}}
\end{figure}

N-body simulations show that the non-linearities cause the power spectrum amplitude to increase relative to the
linear prediction at scales $0.1 < k\, (\text{Mpc}/h) \lesssim 10$ as power is transferred from large scales due to mode coupling. Accounting for this is important for predicting correctly smaller-scale phenomena and thus is implemented in \camb\ using the Halofit routine \cite{Takahashi:2012em}. Halofit is calibrated to replicate the results of $\Lambda$CDM N-body simulations interpolating over a range of $\Lambda$CDM parameters. One should have no expectation that it will work well in an extended scenario such as the one described in this paper. Indeed, simulation of WDM scenarios find that Halofit significantly overestimates the small-scale power spectrum \cite{Viel:2011bk}. Since we are also investigating lower DM masses, this effect is likely to be much more severe.

\begin{figure}
\centering
\vspace{0cm}\rotatebox{0}{\vspace{0cm}\hspace{0cm}\resizebox{0.49\textwidth}{!}{\includegraphics{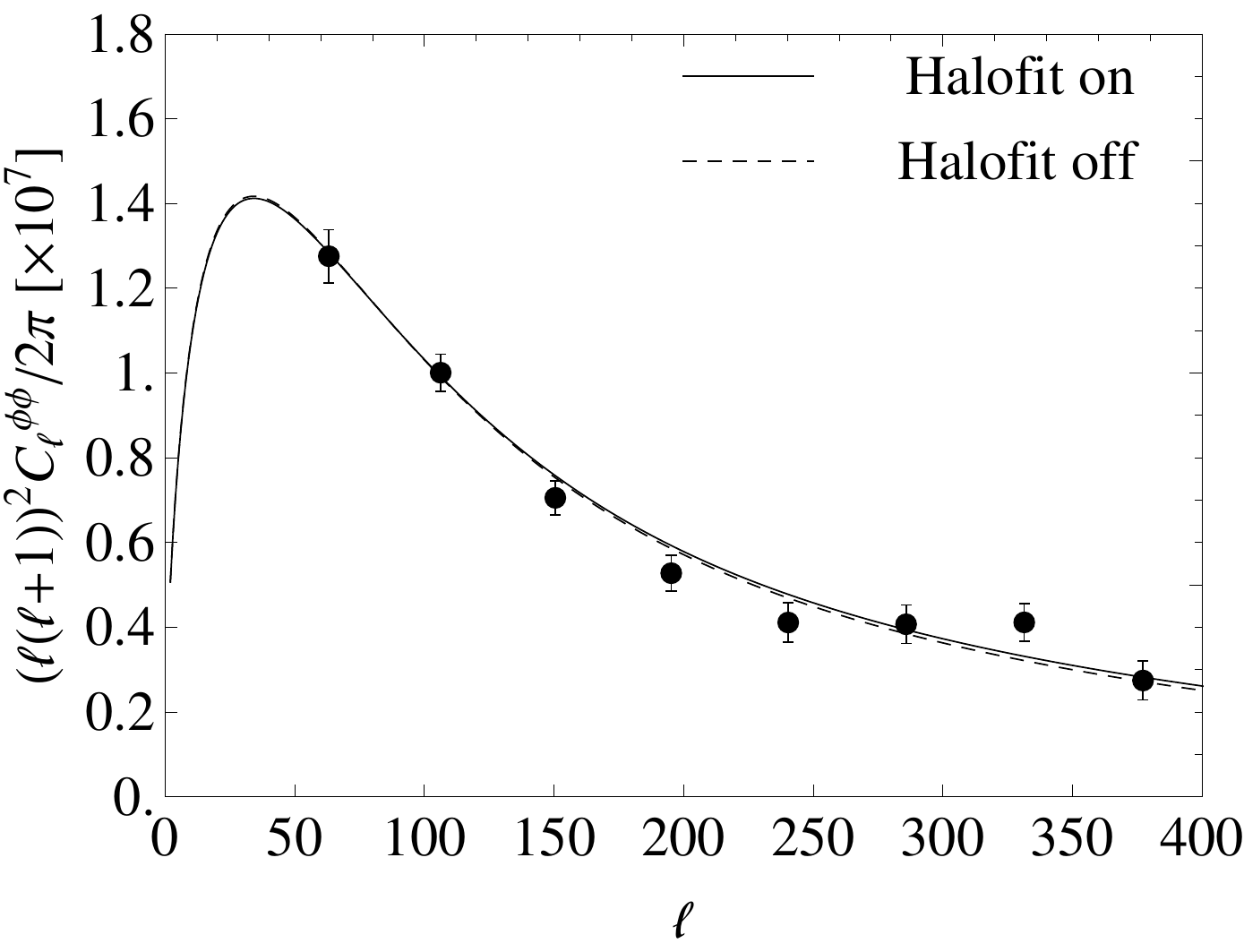}}}
\caption{The lensing potential for $\Lambda$CDM with (black solid line) or without (black dashed line) Halofit. Data points from Planck 2015 derived from the observed trispectrum \cite{Ade:2015zua}. \label{fig:bands}}
\end{figure}

One should thus be very careful with a method like Halofit whenever the correction to the power spectrum is scale-dependent. We have found that keeping Halofit turned on in \camb\ results in posteriors that are highly suspicious: it introduces various oscillations in the posterior parameter probabilities and affects the convergence of the Markov chains. We have thus decided to switch Halofit off in both the calculations of the power spectra and the trispectrum. Since the trispectrum is obtained only from multipoles $\ell<400$, the effect there is not substantial, see Fig. \ref{fig:bands} for a direct comparison. On the other hand, the lack of this correction could bias the CMB lensing constraints from the power spectrum \cite[e.g.\ Fig.~1]{Namikawa:2014xga}. In order to estimate the impact we include the $A_\text{lens}$ parameter with and without Halofit in a $\Lambda$CDM analysis using the Planck power spectra as well as weak lensing data, and find that there is no significant change in $A_\text{lens}$ or any of the other parameters.

Comparing the theoretical predictions with and without Halofit we find that the changes to the CMB power spectrum are at the level of a few per mille, while the lensing power spectrum varies by a few percent for the scales of interest as can be seen in Fig.\ \ref{fig:bands}. Both changes are smaller than the error bars of the data.
Additionally, in the cases where, as mentioned above, using Halofit leads to strange-looking posteriors, we find that the upper limits of the fluid parameters are not very different. We conclude that the data sets used here are sufficiently conservative so that the behavior of the model on non-linear scales is not very important. In what follows, we will therefore always quote the results without a Halofit correction.

\subsection{Initially Relativistic DM}
\label{sec:InitRel}

\begin{figure*}[!t]
\centering\vspace{0cm}\rotatebox{0}{\vspace{0cm}\hspace{0cm}\resizebox{0.99\textwidth}{!}{\includegraphics{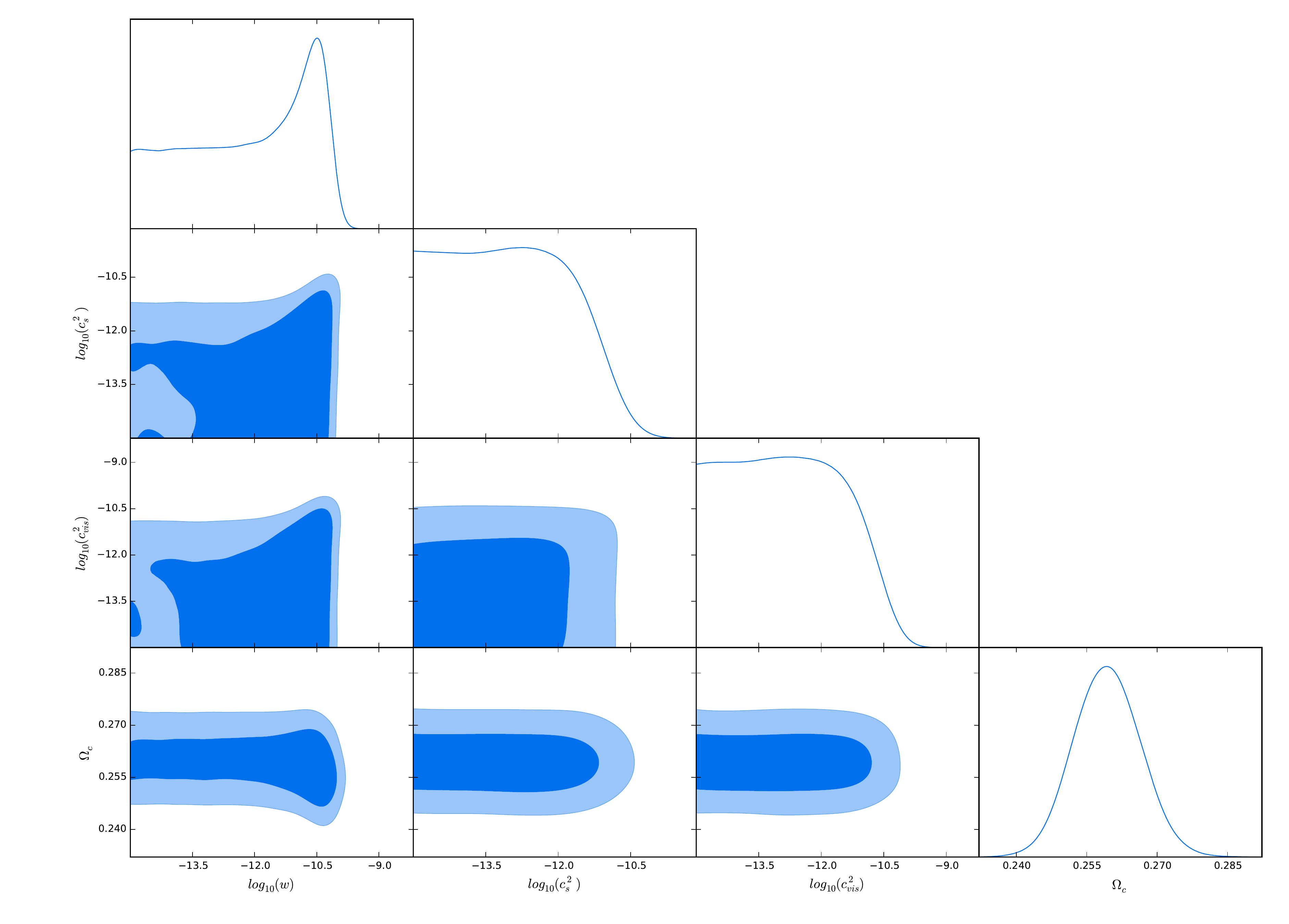}}}
\caption{The 2D $68\%$, $95\%$ confidence contours and the 1D marginalized posterior distributions for the parameters $(w_0,\csz, \cvisz)$ of the initially relativistic model for the Planck (without trispectrum), BAO and SNIa data.  Halofit is turned off. \label{fig:reltest1}}
\end{figure*}

As our headline figures, we choose to report the constraints using the full Planck power spectrum data (including polarization), but excluding the CMB lensing reconstructed from the trispectrum. We also include the distance data from SNIa and BAOs. As mentioned in the previous section, Halofit was switched off. We find that Planck data places upper bounds on the DM parameters: $\log_{10} w_0 < -10.0$, $\log_{10} \csz^2 < -10.7$, $\log_{10} \cvisz^2 < -10.3$ at the $99\%$ confidence level. Interestingly, despite the fact that $w$ and the sound speed affect very different physics, all the bounds are approximately the same. A non-zero $w_0$ provides an insignificantly better fit to Planck data ($\Delta\chi^2=\chi^2-\chi^2_{\Lambda CDM}=-0.6$) for the best fit $\log_{10} w_0 = -10.7$), but the posteriors for $\log_{10} \csz^2$ and $\log_{10} \cvisz^2$ decrease monotonically toward their upper bound. In Fig.~\ref{fig:reltest1} we show the 2D $68\%$, $95\%$ confidence contours and the 1D marginalized posterior distributions for our hydrodynamical parameters $(w_0,\csz^2, \cvisz^2)$.

Such constraints on the DM parameters imply, as expected, that already by recombination the dark matter must be highly non-relativistic $w(z_\text{rec}) \lesssim 10^{-3}$ and similarly for the other parameters. This implies that the observables are mostly affected by the region $x\gtrsim 100$ of the approximation \eqref{eq:paramer-w}, where the deviation from the full numerical solution is negligible. Thus, improving this approximation would have no effect on the constraints. In addition, in this high-$x$ region the difference between fermions and bosons is very small, so that it will be very difficult to tell the two apart based on cosmological large-scale structure data.

We note that the preferred higher values of $w_0$ also allow for a slightly wider range of spectral tilt, although no significant shift occurs in the marginalized posterior. On the other hand, values of the sound speeds close to the upper bound result in slightly lower $\sigma_8= 0.805\pm 0.030$, giving a slightly wider posterior than $\Lambda$CDM for which we find $\sigma_8= 0.830\pm 0.015$. There is no significant effect on $H_0$.

Allowing for a free neutrino-mass-sum parameter does not significantly change the constraints on the DM parameters. However, the constraints on the neutrino masses are weakened, with $\sum m_\nu < 0.35$~eV, compared to the $\Lambda$CDM standard of $\sum m_\nu < 0.23$~eV \cite{Ade:2015xua}. Finally, we find that adding the Planck trispectrum does not significantly change any of the fits.

The independent constraints on the three fluid parameters are compatible with the expected hydrodynamical scenario: $w_0=\csz^2$ and $\cvisz^2=0$. Forcing this scenario, which is not disfavored compared to the fully free one, gives a one-parameter model with an upper bound $\log_{10} w_0 < -10.6$, with no preference for values different from zero. We will use this upper bound to derive constraints on DM particle mass in section~\ref{sec:Implications}.

The ultraconservative cut of the weak-lensing-shear data from CFHTLenS, together with distance measurements, allows for a slightly wider range of extended DM parameters. With $A_S$ and $n_S$ fixed to their $\Lambda$CDM best-fit values, WL allows for $\log_{10} w_0 < -8.1$, $\log_{10} \csz^2 <-8.2$ and $\log_{10} \cvisz^2 <-7.7$. The best fit lies at $\log_{10} \csz^2= -7.2$, but with $\Delta\chi^2 = -1.7$ it is only a marginal improvement over $\Lambda$CDM.  We thus see that this kind of model is not capable of substantially improving the fit to the WL data over concordance.

Furthermore, since the constraints from WL are significantly weaker than from Planck, the combined fit for Planck plus distance probes will be mostly constrained from the Planck-only plus the distances data only. In this case, we find the following upper bounds on the DM parameters: $\log_{10} w_0 < -9.9$, $\log_{10} \csz^2 < -10.5$, $\log_{10} \cvisz^2 < -10.3$ at the $99\%$ confidence level. The corresponding 2D $68\%$, $95\%$ confidence contours and the 1D marginalized posterior distributions can be seen in the appendix in Fig. \ref{fig:reltest3}.

We also note that if the full CFHTLenS data set \cite{Heymans:2013fya} is used instead of the ultraconservative cut, the results are very different. The full data set together with Planck and distances strongly prefer a non-zero equation of state, $\log_{10} w_0 = -10.1 \pm 0.15$ while the sound speeds have the upper bound $\log_{10} \csz^2 < -12.6$, $\log_{10} \cvisz^2 < -12.1$. Such a detection would be incompatible with a hydrodynamical interpretation. Evidently, a contribution present in the full CFHTLenS data is driving an effect which is in tension with the $a^{-2}$ scaling, although surprisingly it prefers lower speeds suggesting that more power is favored.

%----------------------------------------------------------------------------------------------------------

\subsection{Constant $w,\cs^2,\cvis^2$}
\label{sec:constres}

We now consider complementary constraints, with the fluid parameters $w, \cs^2, \cvis^2$ all constant. This gives the maximum value that any of these parameter is allowed to take, and therefore, when combined with the results of section \ref{sec:InitRel}, can help estimate the redshift at which the parameter is constrained most strongly. This parameterization is also sensitive to some late-time effects incompatible with the $a^{-2}$ scaling.

\begin{figure*}[!t]
\centering
\vspace{0cm}\rotatebox{0}{\vspace{0cm}\hspace{0cm}\resizebox{0.90\textwidth}{!}{\includegraphics{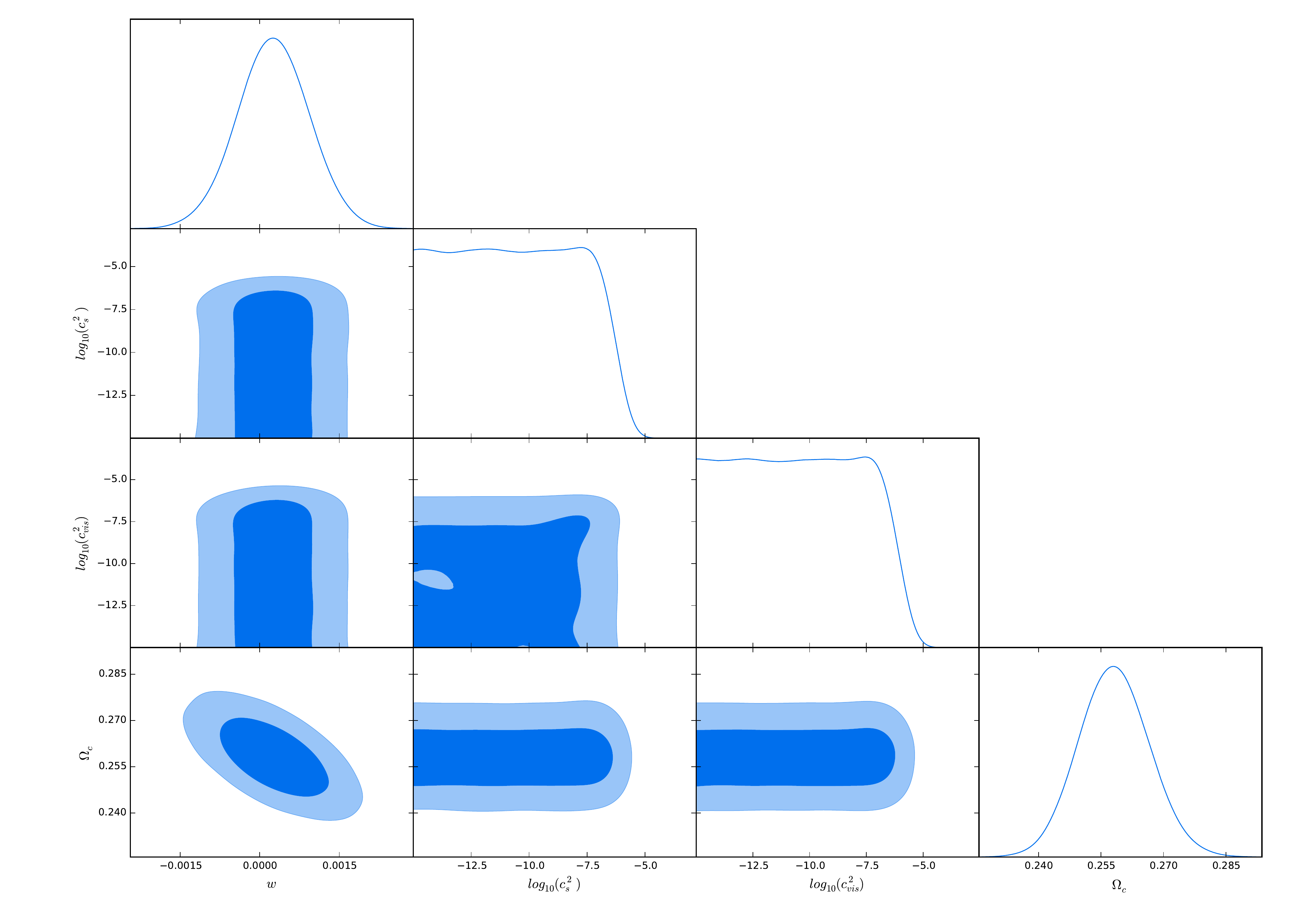}}}
\caption{The 2D $68\%$, $95\%$ confidence contours and the 1D marginalized posterior distributions for the parameters of the model in the case of constant $(w,\cs, \cvis)$ being free to vary, for the Planck (without trispectrum), BAO and SNIa data.
\label{fig:test1c}}
\end{figure*}

We again use the Planck power spectrum data (including polarization), but not the trispectrum, combining it with probes of background geometry from SNIa and BAOs. A constant equation of state for DM is constrained to $w=(-0.26\pm0.68)\cdot 10^{-3}$, i.e.\ no deviation from the standard value of $w=0$ is preferred. Allowing for a non-zero value of $\cs^2$ or $\cvis^2$ does not change the range of allowed $w$. In Fig.~\ref{fig:test1c} we show the 2D $68\%$, $95\%$ confidence contours and the 1D marginalized posterior distributions for the parameters of the model in the case of constant $(w,\cs^2, \cvis^2)$ for the Planck (without trispectrum and Halofit switched off), BAO and SNIa data.

The sound speed parameters $\cs^2$ and $\cvis^2$ are constrained from above, $\log_{10} \cs^2 <-5.9$ and $\log_{10} \cvis^2<-5.7$ at the $99\%$ confidence level. Forcing $w=0$ does not change the upper bounds on the other parameters significantly. We also note that these values are in excellent agreement with a similar analysis performed in ref.~\cite{Thomas:2016iav} and our analysis for a similar model in ref.~\cite{Kunz:2015oqa}.

Comparing the constraints in this scenario with those of section \ref{sec:InitRel} allows us to gain insight into the physics from which the strongest constraints arise. For both constant and initially relatistic $w$, we compare the predicted CMB power spectra for values of $w$ separated by approximately 1$\sigma$. We find that the difference between the two spectra is independent of whether the CMB lensing is included or not and therefore conclude that the effect is generated at large redshifts. Indeed, comparing the constraints for the two parameterizations, $w_0(1+z)^2 \sim w$, gives $z\sim 2000$, reasonably close to $z_\text{rec}$. This results from a non-zero equation of state for DM around recombination changing the angular scale of the CMB peaks, which is extremely well constrained by Planck.

\begin{figure*}[!t]
\centering
\vspace{0cm}\rotatebox{0}{\vspace{0cm}\hspace{0cm}\resizebox{0.975\columnwidth}{!}{\includegraphics{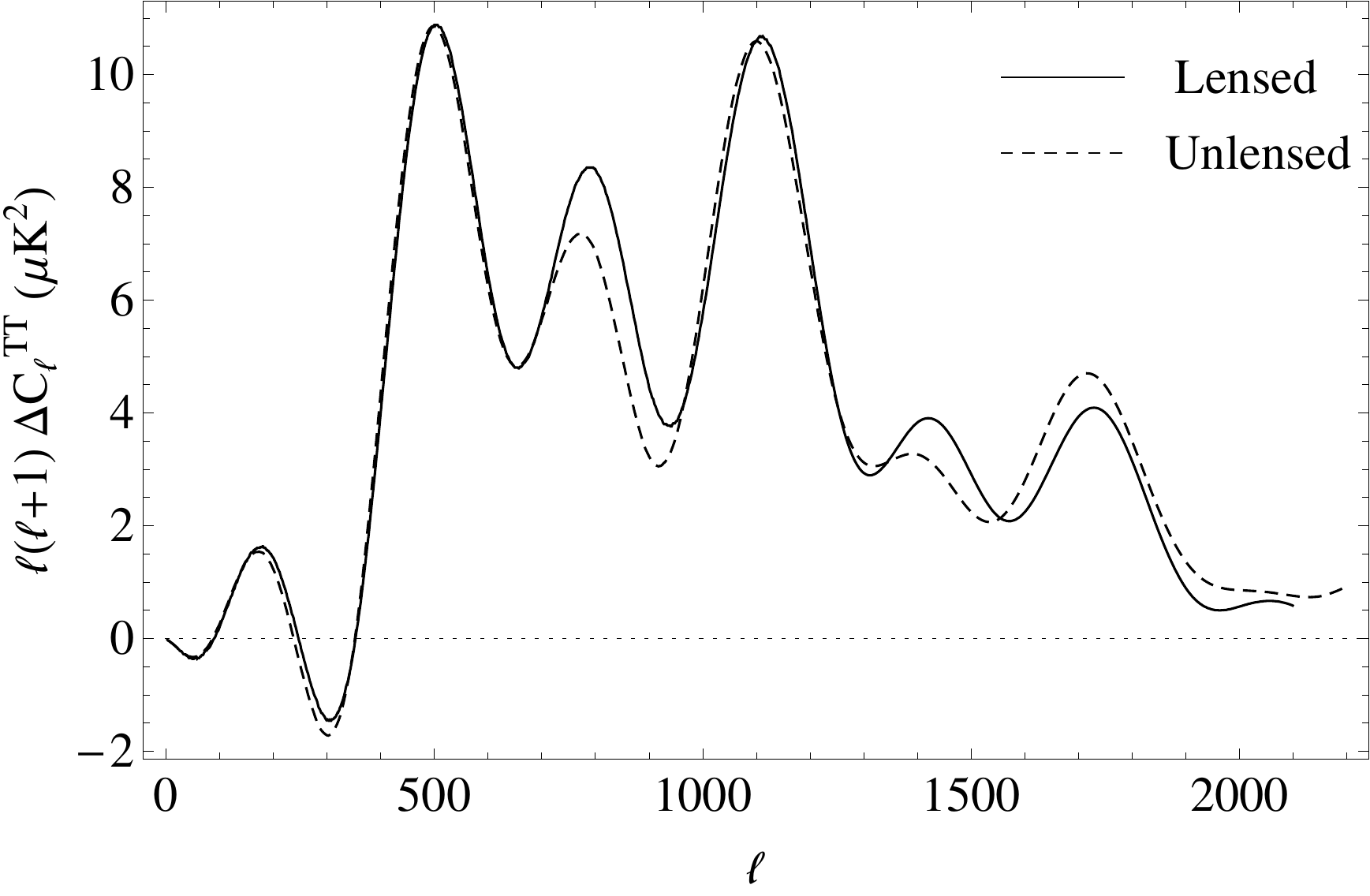}}}
\vspace{0cm}\rotatebox{0}{\vspace{0cm}\hspace{0cm}\resizebox{1.0\columnwidth}{!}{\includegraphics{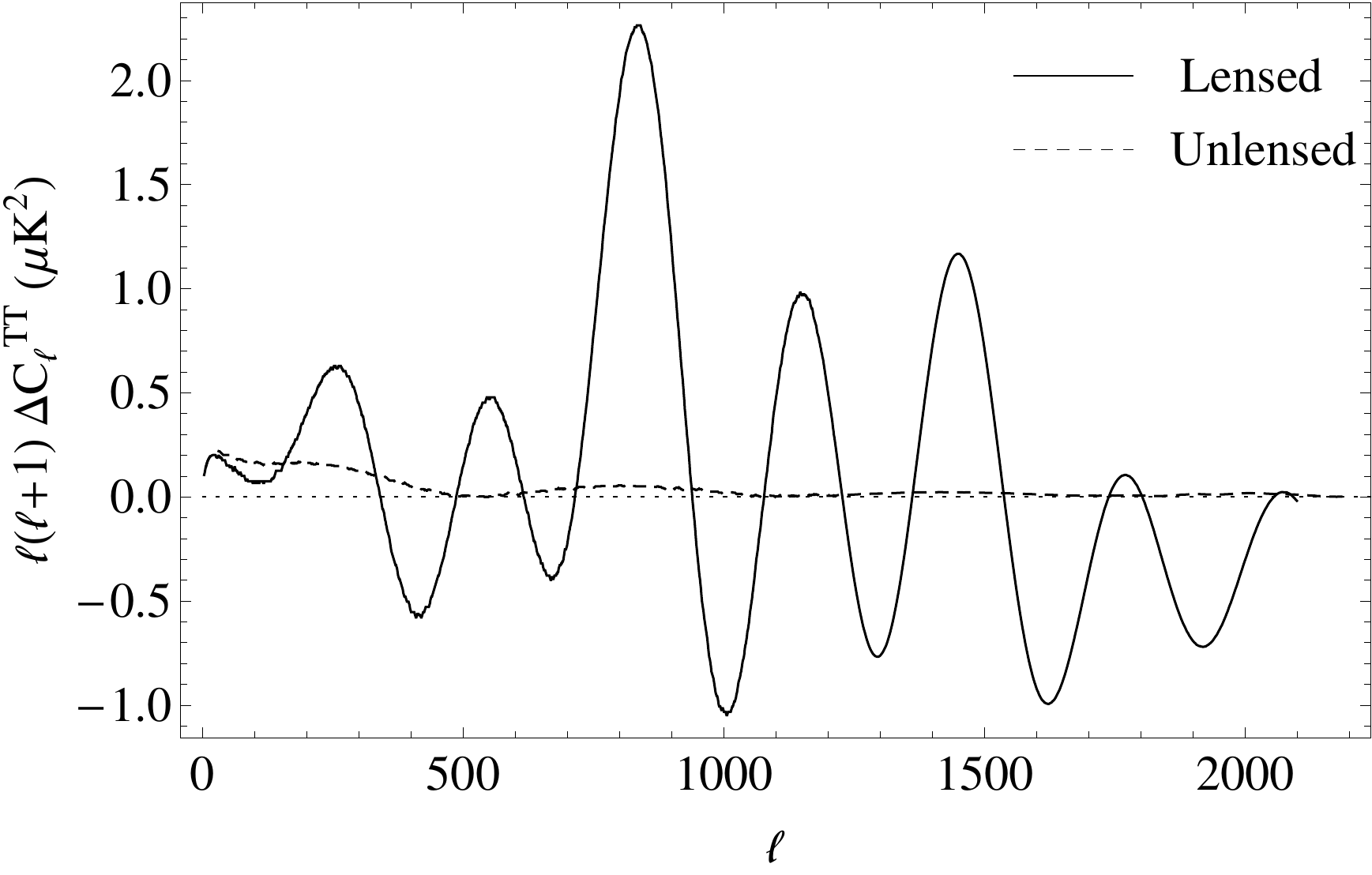}}}
\caption{The difference in the CMB TT powerspectrum for the initially relativistic (left) and constant (right) parameterizations between the bestfit and $1\sigma$ value of the sound-speed $c_{s,0}^2$. \label{fig:LensingOnOff}}
\end{figure*}

On the other hand, when this procedure is repeated for the sound speed, we find that, in the case of constant parameterization, turning off the lensing removes completely the effect from $\cs^2$ on CMB spectra. We can thus conclude that the constraint that $\cs^2<10^{-6}$ comes from low redshifts. Nonetheless, the CMB at recombination is of course sensitive to large values of the sound speed, constraining it to $\cs^2(z_\text{rec})\lesssim10^{-3}$. We infer this from the constraint on the initially relativistic parameterization, by scaling the late-time constraint. We know that this is the right way to look at it because firstly switching lensing on and off in the difference spectra for this case has only very little impact, and secondly because the comparison between the constant and initially relativistic parameterization would imply an effective redshift of $z\approx100$ for the sound speed constraint, which is too high for lensing. We have shown this analysis in figure~\ref{fig:LensingOnOff}.

Allowing the neutrino mass sum $\sum m_\nu$ to vary does not significantly change the constraints on the sound speeds. On the other hand, there is a correlation between larger neutrino mass and higher $w$. Freeing DM properties worsens the constraints on the neutrino masses to $\sum m_\nu < 0.79$~eV.

Just as in the initially relativistic case, WL shear data from CFHTLenS alone, with $A_S$ and $n_S$ held at the $\Lambda$CDM best fit, allows for a larger range of parameters, $w=(-4\pm26)\cdot 10^{-3}$, $\log_{10}\cs^2 <-4.5$, $\log_{10} \cvis^2 < -4.2$. In any case, non-zero values are not preferred significantly. Combining the WL, Planck and distance data sets together leads to a best fit with $w=(-0.12 \pm 0.68)\cdot 10^{-3}$, $\log_{10} \cs^2 <-5.6$ and $\log_{10} \cvis^2 <-5.4$, but with a $\Delta\chi^2= -0.22$ compared to the $\Lambda$CDM case this is not a significant improvement. These higher values sound speed modify the fluctuation amplitude in a scale-dependent manner, yielding $0.67<\sigma_8<0.85$, We show the posteriors in the Appendix, Fig. \ref{fig:test3a}.%
\footnote{Note that the update to the Planck 2015 results has decreased the significance of the improvement to the fit for the model discussed in ref.~\cite{Kunz:2015oqa}.}%

When the full CFHTLenS data \cite{Heymans:2013fya} are combined Planck and distances, the constraints tighten significantly as a result of including much smaller scales. There is no preference for any deviation from the concordance CDM value: $w = (0.68 \pm 0.68)\cdot 10^{-3}$, $\log_{10}\cs^2<-7.8$ and $\log_{10}\cvis^2 < -7.6$.

%----------------------------------------

\subsection{Implications}
\label{sec:Implications}

Our constraint on the sound speed implies that free-streaming scale today has the bound
\begin{equation}
k_\text{FS,0} = \sqrt{\frac{3}{2}}\frac{H_0}{\csz} > 81~h\text{Mpc}^{-1} \,, \label{eq:kfs}
\end{equation}
which scales as $a^{3/2}$ during matter domination.	 This means that despite the scale's being constrained to lie deep in the non-linear regime today, free streaming will have affected much larger scales in the past, erasing the power spectrum also in the linear regime, see Fig.~\ref{fig:halo}. 

Indeed, as demonstrated by figure~\ref{fig:LensingOnOff}, the initially relativistic sound speed is constrained by the CMB power spectrum at recombination and thus implies that the free streaming scale at recombination is $k_\text{FS}(z_\textbf{rec})>1 h$~Mpc$^{-1}$. Only scales smaller than this are allowed to be modified by free streaming for $z>z_\text{rec}$.

On the other hand, the constraint on the constant sound speed is mainly from CMB lensing and thus it implies that the free-streaming scale at low redshifts is constrained to be $k>0.2 $~Mpc$^{-1}$, a much larger scale than implied by eq.~\eqref{eq:kfs}. Recombination thus provides a much better constraint on the DM properties, but it does not take any new physics afterwards into account ($\propto a^{-2}$ scaling). Thus new physics beyond redshifting which might occur after recombination are constrained much more weakly.

Constraints on the power spectrum amplitude from the Lyman-$\alpha$ forest mean that scales at about $k\sim 10$~Mpc$^{-1}$ should also have been unaffected by redshift $z\sim 5$ and  offer an even stronger constraint on the free-streaming scale at that redshift\cite{Viel:2013apy}.

A measurement of the sound speed is equivalent to a measurement of the dispersion (mean-squared particle velocity) $\sigma^2_{v}$ of the dark matter and the constraint implies that today \cite{Shoji:2010hm}
\begin{equation}
\sigma_{v0} = \frac{3}{\sqrt{5}}\csz < 2.0 \, \mathrm{km/s} \,.
\end{equation}
For redshifting collisionless particles this dispersion scales as $a^{-1}$ and this tight constraint is really a result of the limits imposed by the observed recombination physics.

The limits arising from low redshifts resulting from the compatibility of CMB lensing with standard CDM give
\begin{equation}
	\sigma_{v}^\text{late} < 450  \, \mathrm{km/s} \,. \label{eq:latedisper}
\end{equation}
This latter constraint is compatible with the $\sim$300~km s$^{-1}$ expected for the typical peculiar velocities or the dispersion in virialized objects.

If one takes to heart the approach of the effective field theory of large-scale structure (EFTofLSS) proposed in ref.~\cite{Baumann:2010tm}, the shell-crossing and non-linear dynamics at small scales should be describable using effective hydrodynamical corrections to the energy-momentum tensor of the dark matter of sufficient size to contribute already at quasi-linear scales (e.g.\ BAO reduction). Thus according to this approach one should expect to detect at already at quasi-linear scales the influence of the non-linearities through a sound-speed or viscosity speed, with the fiducial size measured from N-body simulations in ref.~\cite{Carrasco:2012cv} of $\cs^2=10^{-6}$ today, a little below the largest sound speeds compatible with CMB lensing data \eqref{eq:latedisper}. An alternative approach of ref.~\cite{Blas:2015tla} can be interpreted as predicting $\cs^2+\cvis^2 \sim 10^{-7}$.%
\footnote{Strictly speaking, both the approaches predict a value for $\cs^2$ that scales approximately as $a$, rather than a constant. We have also run this case and find the upper bound from Planck is a little weaker, $c_\text{s}^2<10^{-5}$.}%

It is thus interesting to note that the full CFHTLenS data constrains the low-redshift value of the sound speed to $\cs^2<10^{-7.5}$ at the 99\% confidence level while also preferring a larger value of $w$ than Planck alone, i.e.\ suggesting that more power is favored. This may well be a result of the fact that the data are marginalized over e.g.\ non-linear intrinsic alignments, removing some of the EFT signal. Moreover, we have not properly incorporated the full structure of the EFTofLSS operators, which may well be biasing our conclusions. Nonetheless, if the EFT approach is valid, we should be \emph{detecting} effective hydrodynamical  corrections at intermediate scales as a good match for the effect of the non-linear physics at short scales. This sort of constraints from data containing the full non-linear information obtained on quasi-linear scales should be able to test the predictivity of the EFTofLSS approach.\\

We now turn to a discussion of what the constraints above imply for fundamental properties of dark matter. The usefulness of the parametrization \eqref{eq:paramer-w} is that, on the assumption that dark matter is collisionless and stable, it is independent of the actual phase-space distribution for the DM. It instead merely exploits the redshifting of the DM momentum. The constraints presented essentially come purely from recombination physics, since the implied particle momenta at late times are much too small for any effect on CMB lensing. Adding in information on the normalization of the shape/amplitude of the matter power spectrum at smaller scales improve these significantly (e.g.\ the Lyman-$\alpha$ forest).

The constraints point toward the standard expectation for non-relativistic collisionless matter: $w\approx \cs^2$ and higher multipoles of the Boltzmann hierarchy are suppressed, $\cvis^2\approx0$. Moreover, they imply a DM that is non-relativistic already by recombination ($w(z_\text{rec})\lesssim 10^{-3}$) and therefore a more precise modelling of the Boltzmann hierarchy is unlikely to significantly change the constraint. We are thus going to use the results assuming $w=\cs^2$ and $\cvis^2=0$ to constrain the DM mass.

\paragraph{Pure Warm Dark Matter}

In Warm Dark Matter scenarios, one typically assumes that the DM froze out while relativistic with a Fermi-Dirac distribution in phase space,
\begin{equation}
	f(\boldsymbol{q}) = \frac{\chi}{e^{q/T_\text{dec}}+1}\,, 		\label{eq:fRel}
\end{equation}
where $q$ is the constant comoving momentum and $T_\text{dec}$ is the temperature at which the DM decoupled. A suppression factor $\chi$ can appear in e.g.\ the case of sterile neutrinos, where the number density is suppressed as a result of the small mixing with active neutrinos \cite{Dolgov:2000ew}. The abundance of such a dark matter is then predicted to be
\begin{equation}
	\Omega_\text{c0}h^2 = \chi \left(\frac{m}{92~\text{eV}}\right) \left(\frac{10.75}{g_*^\text{dec}}\right)
\end{equation}
where $g_*^\text{dec}$ is the number of relativistic species at decoupling \cite{Gershtein:1966gg,Cowsik:1972gh}. Extending the dark-matter models as we have done does not significantly change the constraints on $\Omega_\text{c0}$ or $H_0$ and therefore the above result is a constraint on the required value of $\chi$ for a given relativistic species content and DM mass.

For the distribution \eqref{eq:fRel}, the equation of state today is then given by
\begin{equation}
		w_0 = 4.3 \left(\frac{ T_\gamma}{m}\right)^2\left(\frac{4}{11} \frac{10.75}{g_*^\text{dec}}\right)^{\frac{2}{3}}\,,
\end{equation}
where we have expressed using the current CMB temperature \cite{Colombi:1995ze}. We thus have
\begin{equation}
		m \left(g_*^\text{dec} \right)^{\frac{1}{3}} = 3.3 \frac{T_\gamma}{\sqrt{w_0}}
\end{equation}
which is independent of the suppression factor $\chi$. Thus any cosmological constraint depends only on $w_0$ and $\Omega_\text{c0}$ always leaves one of $g_*^\text{dec}$, $m$ or $\chi$ unfixed.

The CMB temperature today is $T_\gamma = 2.725\, \mathrm{K} = 0.235\, \mathrm{meV}$ \cite{Fixsen:2009ug}. The corresponding mass bound then becomes $m \left(g_*^\text{dec} \right)^{\frac{1}{3}} > 155\, \mathrm{eV}$. If the DM decoupled together with the neutrinos, we have $g_*^\text{dec}=10.75$ and $m> 70$~eV.

This is a much weaker bound than the Tremaine-Gunn bound requiring that $m>400$~eV in order for the gravitational well of galaxies to overcome the Fermi pressure \cite{Tremaine:1979we}. On the other hand, constraints from the Lyman $\alpha$ forest require that $m>3.3$~keV so that the power spectrum remains sufficiently unsuppressed at scales up to $k\sim 10 h$~Mpc$^{-1}$ \cite{Viel:2013apy}. These bounds are  stronger than from CMB alone. Indeed, we can conclude that satisfying these non-CMB bounds leaves any current CMB observables completely unaffected.

\paragraph{Mixed Warm and Cold DM (WCDM)}

Some  models of dark matter predict that in addition to a thermal distribution for a fraction $R$ of the DM, a large fraction $1-R$ of the DM is very cold, with momentum $q\approx0$,
\begin{equation}
		f(\boldsymbol{q}) = (1-R)n_0\delta^{(3)}(\boldsymbol{q}) + \frac{R\chi}{e^{q/T_\text{dec}}+1}	
\end{equation}
where $n_0 = 6\pi\chi\zeta(3)(a_\text{dec} T_\text{dec})^3$ is the number density as would be given by the standard distribution \eqref{eq:fRel}. Such a combined distribution can be a result of a resonant production of sterile neutrinos (e.g.\ \cite{Boyarsky:2008mt}).

The effect of this distribution can be mapped onto our parametrization, with fluid parameters modified as
\begin{equation}
	(w,\cs^2,\cvis^2)_\text{WCDM} \rightarrow (Rw,R\cs^2,R\cvis^2)_\text{WDM} \,.
\end{equation}
Even though the initial conditions we have modeled do not reflect the behavior of such a DM, provided that at recombination $w(z_\text{rec}),\cs^2(z_\text{rec}),\cvis^2(z_\text{rec}))\ll 1/3$, our constraints can be remapped to
\begin{equation}
	m \left(g_*^\text{dec} \right)^{\frac{1}{3}} > 3.3 T_\gamma\sqrt{\frac{R}{w_0}} \,.
\end{equation}
The importance of this rescaling is that all mixed WDM/CDM scenarios have the same effect on observables, provided that the WDM component is non-relativistic already by recombination. Thus no such scenario will offer a solution to the Planck/CFHTLenS tension. Since standard massive neutrinos are relativistic at recombination, such a simple rescaling cannot be used, but rather their behavior in an interpolation between the $a^{-2}$ scaling and the constant parameterization.

This mixed scenario can also be used to describe axion dark matter, where the majority of the axions exists in a condensate with momentum $\boldsymbol{q}=0$, with a small fraction surviving in a thermal distribution. Axions that are light enough have de Broglie wavelengths which are of cosmological size. This gives rise to an effective pressure even when they are in the condensate and can erase structure at small scales \cite{Hu:2000ke}. This has a similar effect to the one described , and the lack of observed deficit of power in the CMB prevents axions with masses $m<10^{-25}$~eV from comprising the majority of the dark matter \cite{Hlozek:2014lca}.

\paragraph{Freezeout while Non-Relativistic}
A cold relic freezes out when non-relativistic and therefore has a Maxwell-Boltzmann distribution
\begin{equation}
	f(\boldsymbol{q}) = \frac{g}{(2\pi)^3} e^{\frac{-q^2}{2mT_\text{dec}}}, \label{eq:MBf}
\end{equation}
where $g$ is the number of states. Strictly speaking, considering cold relics is not compatible with our initial conditions, since we assume that decoupling has already occurred while the species are relativistic. However, since the posterior is insensitive to the initial conditions, the error thus generated is not significant.

Integrating over the distribution \eqref{eq:MBf} we obtain
\begin{equation}
	w = \frac{T_\text{dec}}{m} \left(\frac{a_\text{dec}}{a}\right)^2 \,,
\end{equation}
noting that we have a linear dependence on the decoupling temperature but still a quadratic one on the scale factor. We can replace the dependence on $a_\text{dec}$ and $T_\text{dec}$ with the CMB temperature and $x_\mathrm{f} \equiv m/T_\text{dec}$,
\begin{equation}
	w_0 = \frac{T_\gamma^2 x_\mathrm{f}}{m^2} \left(\frac{4}{11} \frac{10.75}{g_*^\text{dec}} \right)^{\frac{2}{3}} \, ,
\end{equation}
yielding the result
\begin{equation}
\frac{m}{\sqrt{x_\mathrm{f}}} \left(g_*^\text{dec}\right)^{1/3} = 1.6 \frac{T_\gamma}{\sqrt{w_0}}\,. \label{eq:cold1}
\end{equation}

In principle, there are two distinct freezeout conditions: the usual chemical freezeout, which sets the final abundance of the DM and a kinetic freezeout which determines when the DM stops interacting with other species, e.g.\ the photons.

%The chemical freezeout condition $x_\mathrm{f}$ and the final DM abundance are given by the approximate solutions (ref.~\cite[eq. (5.44,5.47)]{KolbTurner},
\begin{align}
	x_\mathrm{f} &\approx -0.3 + \ln \left(m_1\sigma_{27} / \sqrt{g_*^\text{dec}} \right) \label{eq:xf} \,,\\
	\Omega_\text{c} h^2 &= 1.05\frac{(g_*^\text{dec})^{1/2}}{g_{*\text{S}}^\text{dec}} x_\text{f}\sigma_{27} \,, \notag
\end{align}
where we have defined the convenient dimensionless averaged annihilation cross-section and mass in units of eV,
\begin{equation}
 \sigma_{27} \equiv \frac{ \left< \sigma_\text{A} |v|\right>}{10^{-27}~\text{cm}^3\text{s}^{-1}}\,, \qquad m_1 \equiv \frac{m}{1\text{ eV}} \,.
\end{equation}
and we have assumed that the annihilations proceed through an $s$-wave. On the other hand, following ref.~\cite{Chen:2001jz}, we can estimate the scattering cross-section of the cold DM with neutrinos/photons as
\begin{equation}
	\sigma_{\text{scatt} }\sim \left(\frac{T}{m}\right)^4 \sigma_\text{A} \,.
\end{equation}
where $T$ is the typical energy of the photons in the universe at the time. The chemical freezeout occurs when the scattering rate $\Gamma=n_\gamma \sigma_\text{scatt}\sim H$, which can be rewritten in our chosen units as
\begin{equation}
\left(\frac{m}{T}\right)^4 \sim \frac{m_1 \sigma_{27}}{100\sqrt{g_*^\text{dec}} }\,.
\end{equation}
For the sort of DM masses that we are constraining, the chemical freezeout occurs at the same time or later as the kinetic freezeout. Only when the mass reaches $m\sim 500$~keV does this estimate imply that the kinetic freezeout is delayed compared to the chemical one. For the purpose of this analysis, we will neglect the kinetic freezeout henceforth.
\begin{figure}[!t]
	\includegraphics[width=1\columnwidth]{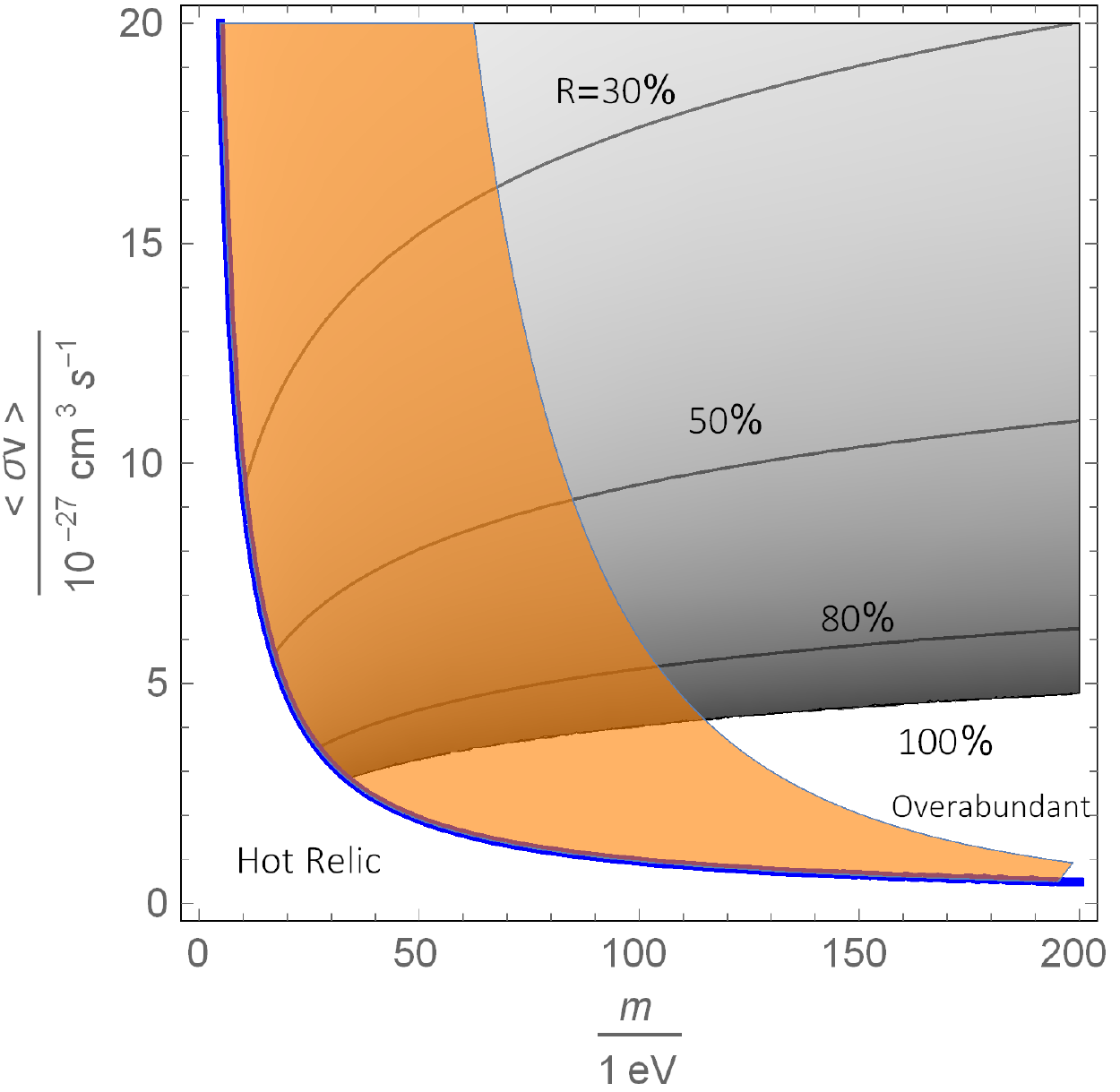}
	\caption{Constraints on thermal cold relic DM.	Only the region to the right of the blue line ($x_\text{f}=3$) corresponds to freeze out while non-relativistic. The region shaded in gray is allowed with our dark matter species contributing an increasing fraction $R$ of the total DM for smaller cross-sections, with the lower boundary corresponding to this DM's comprising all of dark matter. The excluded orange region is the result of the analysis presented in this paper and its translation to DM properties through the result \eqref{eq:cold1}.\label{fig:ColdRelics}}
\end{figure}

We thus take as the boundary of non-relativistic freeze-out the condition $x_\text{f}\gtrsim 3$. In figure~\ref{fig:ColdRelics}, we show that our constraints obtained from the CMB restrict the parameter space for cold thermal relics in the $m\sim 10-100$~eV mass range. Such DM species would freeze out between Big-Bang nucleosynthesis (BBN) and recombination (and therefore have $g_*^\text{dec}=3.36$) and contribute as a relativistic degree of freedom during BBN. This is in tension with the data, but marginally allowed \cite{Ade:2015xua}. On the assumption that this DM contributes the totality of the DM, we obtain the constraint that
\begin{equation}
	m > 104~\text{eV} \,. \label{eq:coldmass}
\end{equation}
Although very low compared to the typical scenarios where $m\sim 100$~GeV, such low masses are not incompatible with technical requirements: the freeze-out occurs before recombination, the mass of the mediator is much larger than $m$ and yet non-relativistic during BBN. Yet again, we must stress that the Tremaine-Gunn bound of $m>400$~eV for fermions remains stronger \cite{Tremaine:1979we}.

The constraints from the Bullet cluster \cite{Markevitch:2003at} do not restrict the parameter space in this region and the lightness of the DM would mean that it cannot decay to leptons, but only to photons, making such models compatible with the heating of the intergalactic medium \cite{Cirelli:2009bb}. Models with such masses would produce a line in the X-ray spectrum. This would be swamped by the emission of the hot gas in clusters and thus is not observable for masses $m\sim 100$~eV \cite[Fig.~19]{Adhikari:2016bei}. The constraint \eqref{eq:coldmass} implies that the freezeout took place at $T_\text{dec}>450$~eV. This constraint is on the boundary of sensitivity of the CMB spectrum to energy injections through $\mu$ distortions of the CMB spectrum \cite{Hu:1992dc}. This constraint is also complementary to the those obtained from the $\mu$ distortions caused by scattering of photons and nucleons off DM particles prior to recombination. Such a constraint on the mass can be much stronger ($m> 0.1$~MeV) provided that there is a sufficiently large coupling between the DM and baryons/photons \cite{Ali-Haimoud:2015pwa}.

\section{Summary and Discussion}
\label{sec:discussion}

The Planck mission has provided us with an unprecedented quality of data for the CMB, which not only is sensitive to the universe at recombination, but is precise enough to see the effect on the propagation of the CMB photons of the gravitational field through CMB lensing. We have used these data to constrain the hydrodynamical parameters of the dark matter under the very general assumption that it be collisionless and therefore the momenta redshift with the scale factor. This leads to the constraint that today's value of the equation of state $\log_{10} w_0 <-10.0$, the sound speed $\log_{10} \csz^2 < -10.7$ and the viscosity parameter $\log_{10} \cvisz^2 <-10.3$, based on the assumption that these parameters scale as $a^{-2}$ after they exit their relativistic behavior. The rough equality of all these parameters implies that there is no evidence for any unexpected non-hydrodynamical corrections in the evolution of the DM energy-momentum tensor.

The constraints arise from different physics: the constraint on $w$ is mainly from the correction it would introduce to the expansion rate during recombination, which is limited to $w(z_\text{rec})<10^{-3}$. The strongest direct constraint on the sound speed, on the other hand, comes from CMB lensing at low redshifts, $c_s^2 < 10^{-6}$. Nonetheless there is also a constraint from $z_\text{rec}$ requiring that $c_s^2(z_\text{rec}) \lesssim 10^{-3}$. Owing to the $a^{-2}$ scaling, the recombination constraint dominates over the lensing constraint today.

These constraints are largely independent of the particle model of dark matter and its production mechanism, depending only on the conservation of the phase-space distribution function and therefore the translation to parameter's values today is very general. If the dark matter comprises multiple species then these constraints apply to the density-weighted average under the assumption that all the subcomponents are non-relativistic by recombination.

On the other hand, the translation into a constraint on particle properties for DM is model dependent. For example, for warm dark matter which froze out while relativistic, these constraints can be translated to a constraint on a combination of the mass and the number of relativistic species at decoupling, $m (g_\text{dec})^{1/3}>155$~eV; the constraint on mass of thermal cold relics is of similar magnitude, $m> 104$~eV.  We have also shown how such a constraint can be easily translated into one for a a model with more than one dark matter species.

These constraints are of course much weaker than those provided by Lyman-$\alpha$ forest observations: $m>3.3$~keV implies that $\log_{10} \csz^2 < -14$, i.e.\ a constraint on the sound speed squared better by nearly three orders of magnitude \cite{Viel:2013apy}. Nonetheless, to obtain the constraints given here, we are using purely linear physics and thus they are very robust. Lyman-$\alpha$ results depend on understanding the ionization history of hydrogen, which requires the detailed modelling of its hydrodynamics and an understanding of the thermal history of the intergalactic medium \cite{Garzilli:2015iwa}. The generality of the scaling does imply that any measurements of the amplitude or shape of the fluctuation power spectrum, provided they are made at scales which have not undergone shell crossing, can push the constraint on the mass much further.

Our results have an important implication for attempts to decrease the amplitude of fluctuations at smaller scales: The predictions for cosmological observables for a very large class of  DM models are the same (up to some remapping of the particle properties). We find that they do not improve the fit, and even if they did, they would be excluded by Lyman-$\alpha$ constraints. This is a very general statement: essentially no model in which the underlying particles are collisionless, non-relativistic and redshifting can achieve an improved fit, since it will always produce the same profile of modification to the power spectrum.  This all results from the very general $a^{-1}$ scaling for the sound speed and therefore the velocity dispersion. The sound speed grows too quickly with redshift, erasing too large a range of scales to allow such a model to be compatible with observations.

In the effective-field-theory approach to large-scale structure, non-linear evolution on small scales  manifests itself on quasi-linear scales through effective hydrodynamical corrections to the DM EMT. Thus at late times one should expect to find that fully non-linear evolution of dark matter can be interpreted as an effective fluid on intermediate scales. The best constraints at late times are provided by CMB lensing, $\cs^2 < 10^{-5.9}$ and are on the margin of the predicted values of ref.~\cite{Carrasco:2012cv}. However, when the full CFHTLenS data are included, this constraint becomes much stronger, $\cs^2 < 10^{-7.5}$. We did not properly model the behavior of the effective DM EMT, the scaling of the effective sound speeds, nor do we have access to lensing data with no attempt to remove effects of non-linearities. A more detailed and exact analysis of this kind of data should lead to a detection of sound speeds of order $10^{-6}$ if the EFTofLSS approach is right. 

Finally, measurements of the amplitude of fluctuations at smaller scales can be very informative as to the fundamental nature of the dark sector, especially if evidence of tension within $\Lambda$CDM remains. This is likely to continue to be a fruitful area of research in the near future.

%----------------------------------------------------------------------------------------------------------

\begin{acknowledgments}
\emph{Acknowledgements.}  We are grateful to R.~Durrer, C.~Germani, E.~Komatsu and O.~Pujolás for fruitful discussions and comments. M.K.\ acknowledges funding by the Swiss National Science Foundation. S.N.\ is supported by the Research Project of the Spanish MINECO, FPA2013-47986-03-3P, the Centro de Excelencia Severo Ochoa Program SEV-2012-0249 and the Ramón y Cajal programme through the grant RYC-2014-15843. I.S.\ is supported by the Maria Sklodowska-Curie Intra-European Fellowship Project ``DRKFRCS''. The numerical computations for our analysis were performed on the Baobab cluster at the University of Geneva. I.S.\ and M.K.\ thank the Galileo Galilei Institute for hospitality during the final stages of this project.

The development of Planck was supported by: ESA; CNES and CNRS/INSU- IN2P3-INP (France); ASI, CNR, and INAF (Italy); NASA and DoE (USA); STFC and UKSA (UK); CSIC, MICINN and JA (Spain); Tekes, AoF and CSC (Finland); DLR and MPG (Germany); CSA (Canada); DTU Space (Denmark); SER/SSO (Switzerland); RCN (Norway); SFI (Ireland); FCT/MCTES (Portugal). A description of the Planck Collaboration and a list of its members, including the technical or scientific activities in which they have been involved, can be found at \url{http://www.cosmos.esa.int/web/planck/planck-collaboration}.
\end{acknowledgments}

\appendix

\section{Boltzmann Hierarchy}
\label{sec:Boltzmann}

Following ref.~\cite{Shoji:2010hm}, a particle ensemble is described by its distribution function in phase space $f(\boldsymbol{x},\boldsymbol{q}, t)$, where $\boldsymbol{q}$ is the momentum conjugate to the coordinate $\boldsymbol{x}$. We transform $\boldsymbol{x}$ to Fourier space and rewrite the conjugate momentum $\boldsymbol{q}$ in terms of a direction $\bhn$ and a comoving momentum magnitude $q$.%
\footnote{See ref.~\cite{Ma:1995ey} for a discussion on the difference between the canonical momentum conjugate to $\bx$ and the comoving momentum; at background level, they are identical.}%
Additionally, we  assume that we can write
this one-particle distribution in terms of a background value and small perturbations, to obtain
\[
f(\bk, \bhn, q, t) = f_0(q,t) \left( 1 + \Psi(\bk,\bhn,q,t) \right) \, .
\]
Here $f_0$ is the unperturbed `background' distribution function which is independent of position and velocity direction due to homogeneity and isotropy. This one-particle distribution function, when in thermal equilibrium has the form
\[
f_0(p,t) = \frac{g}{(2\pi)^3} \left[ e^{\ve(p,t)/aT}\pm 1 \right]^{-1} \, \label{eq:f0}.
\]
with $p$ the proper momentum of the particle as given by the background comoving observer, $g$ the number of spin states and the $+$ for fermions and $-$ for bosons. The comoving energy  of the particle is defined for convenience $\ve(p,t) \equiv a\sqrt{p^2+m^2}$, and the extra factor $a$ cancels appropriately in eq.~\eqref{eq:f0}, meaning there is no explicit dependence on $a$.

At temperature $T_\text{dec}$, the DM interactions freeze out. From this point on, the only evolution is the redshifting of the individual particle proper momenta $p$, but no rescattering to rethermalize at a new temperature is possible. This relates $f_0$ at different times after decoupling
\begin{equation}
f_0(p,a(t_1)) = f_0\left( p \frac{a(t_1)}{a(t_2)},a(t_2)\right) \,,
\end{equation}
which in turn implies that $f_0$ does not evolve as a function of the comoving momentum $q\equiv p/a$. Thus the meaning of the freezeout is to create a distribution function,
\begin{equation}
f_0(q,t) = f_0(q) = \frac{g}{(2\pi)^3} \left[ e^{\sqrt{q^2+m^2}/T_\text{dec}}\pm 1 \right]^{-1} \, .
\end{equation}
It is important to stress that neither $q$ nor the temperature $T_\text{dec}$ are evolving here, but rather are fixed (notice there is no scaling with $a$ even for the mass term). On the other hand, the comoving energy of each individual particle evolves in the standard manner as $\ve = \sqrt{q^2 + a^2 m^2}$.\\

In linear perturbation theory, the
evolution of the perturbations $\Psi$ for a particle species that has decoupled
are given by the linearized, collisionless Boltzmann equation (e.g.\ \cite{Ma:1995ey})
\[
\dd_t \Psi + i \frac{q}{\ve(q,t)} (\bk\bhn) \Psi + \frac{d \ln f_0(q)}{d \ln q} \left[ \dot{\phi} - i \frac{q}{\ve(q,t)} (\bk\bhn) \psi \right] = 0 \, .
\]
This is usually expanded in terms of Legendre polynomials,
\[
\Psi(\bk,\bhn,q,t) = \sum_{\ell=0}^\infty (-i)^\ell (2\ell+1) \Psi_\ell(k,q,t) P_\ell(\mu) \, ,
\]
and written
as a system of coupled ordinary differential equations (the `Boltzmann hierarchy'),
e.g.\ following  \cite{Shoji:2010hm} and using their definitions
\begin{eqnarray}
\Psi'_0(k,q,x) &=& -\frac{q}{\ve(q,x)} \Psi_\ell(k,q,x) - \phi'(k,x) \, , \\
\Psi'_1(k,q,x) &=& \frac{q}{3\ve(q,x)} \left[ \Psi_0(k,q,x)-2 \Psi_2(k,q,x)\right] \nonumber \\
&& -\frac{\ve(q,x)}{3q} \psi(k,x) \, , \\
\Psi'_\ell(k,q,x) &=& \frac{q}{(2\ell+1) \ve(q,x)} \Large[ \ell \Psi_{\ell-1}(k,q,x) \nonumber \\
&& - (\ell+1) \Psi_{\ell+1}(k,q,x) \Large] \, \, \,  (\mathrm{for}~\ell\geq2) \, .
\end{eqnarray}
with $x\equiv kt$. We can see that each higher element of the hierarchy contains an extra factor $q/\ve$. For neutrinos and
other relativistic species $\ve \approx q$ and the whole hierarchy is important. For non-relativistic species on the other hand, $q/\ve \propto a^{-1}$: the higher multipoles are suppressed by powers of $q/\ve$ and this suppression increases with time.

In principle, we can now compute $w$, $c_s^2$ and the anisotropic stress $\sigma$ in terms of $q/\ve$ by solving the
Boltzmann hierarchy and evaluating the appropriate integrals.

\subsubsection{Background equation of state\label{sec:wimp_w}}

For $w$ we work on the level of the background distribution function $f_0$, the average number density,
energy density and pressure are given by
\begin{eqnarray}
\bar{n}(a) a^3 &\propto& \int dq q^2 f_0(q,a) \, , \\
\bar{\rho}(a) a^4 &\propto& \int dq q^2 \ve(q,a) f_0(q,a) \, , \label{eq:rhobg} \\
\bar{p}(a) a^4 &\propto& \int dq q^2 \frac{q^2}{3 \ve(q,a)} f_0(q,a) \, , \label{eq:pbg}
\end{eqnarray}
where we have neglected common prefactors. As discussed above, at decoupling the function
$f_0$ freezes, $f_0(q,a)=f_0(q)$ and only the comoving energies $\ve$ continue to redshift.

If the particles are relativistic then $q \approx \ve$ and thus $\bar{p} = \bar{\rho}/3$.
As particles become non-relativistic, $\ve \approx a m$, with corrections of the order
of $q^2/(a^2 m^2)$. In this case
\begin{eqnarray}
\bar{\rho}(a) a^3 &\propto& m \int dq q^2 f_0(q) = m \bar{n} (a) \, , \\
\bar{p}(a) a^5 &\propto& \int dq q^4 f_0(q) \, .
\end{eqnarray}
The momentum
integration will just give numbers \emph{irrespective} of the form of $f_0$, and so completely generically the equation of state
of non-relativistic particles evolves as
\[
w(a) = \frac{\bar{p}}{\bar{\rho}} \propto \frac{1}{a^2} \, .
\]
Substituting a particular choice of $f_0(q)$ gives a concrete prediction for $w$. We discuss examples in section~\ref{sec:Implications}.

% ----

\subsubsection{Sound speed and viscosity parameter}

The pressure and density perturbations are given by
\begin{eqnarray}
\delta\rho(k,a) &=& \frac{4\pi}{a^4} \int dq q^2 \ve(q,a) f_0(q,a) \Psi_0(k,q,a) \, , \\
\delta p(k,a) &=& \frac{4\pi}{a^4} \int dq q^2 \frac{q^2}{3 \ve(q,a)} f_0(q,a) \Psi_0(k,q,a) \, .
\end{eqnarray}
We see that this is analogous to the situation in Eq.\ (\ref{eq:rhobg}) and (\ref{eq:pbg}),
except that the integral now additionally contains the perturbation $\Psi_0(k,q,a)$.
From the study presented in \cite{Shoji:2010hm} we can see that $\Psi_0(k,q,a)$ is not a strong function of $q$ and therefore can be taken out of the integral. Because of this it is again the case
that $c_s^2(a) \approx w \propto 1/a^2$ when non-relativistic.\\

The velocity potential $\theta$ and the anisotropic stress $\sigma$ are given by
integrals over $\Psi_1$ and $\Psi_2$ respectively. These higher multipoles
$\Psi_\ell$ are suppressed by additional factors $q/\ve$ when the DM is non-relativistic. Thus the anisotropic stress should decay more quickly than the pressure. Since we are truncating the Boltzmann hierarchy and parameterizing the higher multipoles using $\cvis^2$ through eq.~\eqref{eq:sigmaEvol} as a proxy for all the higher contributions, we are unable to model the evolution precisely. Rather, we are interested in testing to what extent there is any evidence for such contributions. We thus choose to parameterize $\cvis^2$ in the same manner as $\cs^2$ to see if there is any evidence for anisotropic stress larger than the pressure. In this case we would need to use the full hierarchy.

\section{Initial Conditions\label{app:IC}}

In what follows, we will use the subscript $\nu$ to denote a quantity describing the relativistic neutrinos, $\gamma$ --- the photons, and $c$ --- the dark matter. Since $w$ is constant when the initial conditions are set, $c_\text{a}^2=w$.

\subsubsection{Constant Parametrization}
When dark matter is subdominant, ref.~\cite{Ma:1995ey} shows that the gravitational potentials are driven purely by the radiation and neutrinos and therefore we have
\begin{align}
	h & =C(k\tau)^2 \,,\\
	\eta &= 2C-\frac{5+4\Omega_\nu}{6(15+4\Omega_\nu)}C(k\tau)^{2}\,, \notag
\end{align}
where $\Omega_\nu\equiv \rho_\nu/(\rho_\gamma+\rho_\nu)$ and $C$ the amplitude for the mode arising from the inflationary initial conditions. This solution is only valid during radiation domination and on superhorizon scales, $k\tau\ll 1$. The initial adiabatic density perturbations are given by
\begin{equation}
\delta_{\gamma}=\delta_{\nu}=-\frac{2}{3}C(k\tau)^{2}\,,
\end{equation}
while the velocity divergences are given by
\begin{equation}
	 \theta_{\gamma}=-\frac{C}{18}k^{4}\tau^{3}\,,\qquad\theta_{\nu}=-\frac{C}{18}\frac{23+4\Omega_\nu}{15+4\Omega_\nu}k^{4}\tau^{3} \,,
\end{equation}
and the anisotropic stress
\begin{equation}
\sigma_{\gamma}=0\,,\qquad\sigma_{\nu}=\frac{4}{3}\frac{C(k\tau)^{2}}{15+4\Omega_\nu} \,.
\end{equation}
In the presence of the gravitational field being driven by these two collapsing relativistic species, the superhorizon evolution of general dark matter follows the following attractor
\begin{align}
\delta_c&=-\frac{(1+w)C(k\tau)^{2}}{2(4+3\cs^2-6w)}\times  \label{eq:NRDMICs}\\
&\qquad\qquad\times\Big((4-3\cs^{2})-\frac{48}{15+4\Omega_\nu}\frac{\cvis^{2}}{1+w}(\cs^{2}-w)\Big)\,,\notag\\
\theta_c&=-\frac{Ck^{4}\tau^{3}}{2(4+3\cs^{2}-6w)}\times \notag\\
&\qquad\qquad\times\Big(\cs^{2}+\frac{16}{3(15+4\Omega_\nu)}\frac{\cvis^{2}}{1+w}\left(2+3\cs^{2}-3w\right)\Big) \,,\notag\\
\sigma_c&=\frac{16C(k\tau)^{2}}{3(15+4\Omega_\nu)}\frac{\cvis^{2}}{1+w}\, , \notag
\end{align}
where again we stress that $w,\cs,\cvis$ are all constant and $w<\frac{1}{3}$. We can also see that in the limit $w=\cs^2=\cvis^2=\frac{1}{3}$, solution~\eqref{eq:NRDMICs} reduces to the superhorizon solution for the neutrinos given above, which was the original motivation for the form of the parametrization~\eqref{eq:sigmaEvol}.

\subsubsection{Initially Relativistic DM}
\label{sec:IC-REL}
Since attractor solutions exist only when there is a single timescale $H^{-1}$, the initial conditions must be set up in the relativistic regime whenever a time-varying equation of state for DM is considered. We thus compute the initial conditions with $w=\cs^2=\frac{1}{3}$ and consistently include the effect of the DM on the gravitational field. Despite the fact that $\cvis^2$ is a phenomenological parameter, we will also give it the standard initial value of $\frac{1}{3}$, showing below that this particular value replicates the expected superhorizon behavior of a relativistic species.

The first implication of this is that the early universe should be considered to consist of three dominant species, photons $\gamma$, relativistic neutrinos $\nu$ and the dark matter $c$. We define the energy density fractions $\Omega_i$ in the usual manner with
\begin{equation}
		\Omega_\gamma = 1 - \Omega_\nu -\Omega_c	\,.
\end{equation}

The parametrization \eqref{eq:sigmaEvol} is constructed so that in the limit $w=\cs^2=\cvis^2=\frac{1}{3}$, the superhorizon solution for the DM is the same as that for relativistic neutrinos, despite the fact that the superhorizon evolution equation for the second moment of the neutrino distribution is not the same \cite[Eq.~(92)]{Ma:1995ey},
\begin{equation}
	\dot{\sigma}_\nu = \frac{2}{15}(2\theta_\nu + \dot{h}+6\dot{\eta}) \,.
\end{equation}
Despite this difference, the superhorizon attractor is modified in the expected manner, with the replacement of $\Omega_\nu \rightarrow \Omega_\nu + \Omega_c$:
\begin{align}
h & =C(k\tau)^2 \,,\\
\eta &= 2C-\frac{5+4(\Omega_\nu+\Omega_c)}{6(15+4(\Omega_\nu+\Omega_c))}C(k\tau)^{2}\,. \notag
\end{align}
The initial density perturbation are adiabatic and therefore all equal
\begin{equation}
\delta_c = \delta_{\gamma}=\delta_{\nu}=-\frac{2}{3}C(k\tau)^{2}\,.
\end{equation}
The photons, as in the standard solution of ref.~\cite{Ma:1995ey}, carry no anisotropic stress and are not affected at all, while the attractor solution for the relativistic neutrinos and relativistic DM becomes
\begin{align}
\theta_c&=\theta_{\nu}=-\frac{C}{18}\frac{23+4(\Omega_\nu+\Omega_c)}{15+4(\Omega_\nu+\Omega_c)}k^{4}\tau^{3} \,,\\
\sigma_c&=\sigma_{\nu}=\frac{4}{3}\frac{C(k\tau)^{2}}{15+4(\Omega_\nu+\Omega_c)} \,.\notag
\end{align}
We stress that the identical superhorizon behavior of the relativistic DM and neutrinos is a constructed coincidence which only occurs for $\cvis^2=\frac{1}{3}$. We also remind the reader that these initial conditions assume that the DM is decoupled and does not exchange energy with other species.

\begin{figure*}[!t]
\centering
\vspace{0cm}\rotatebox{0}{\vspace{0cm}\hspace{0cm}\resizebox{1\textwidth}{!}{\includegraphics{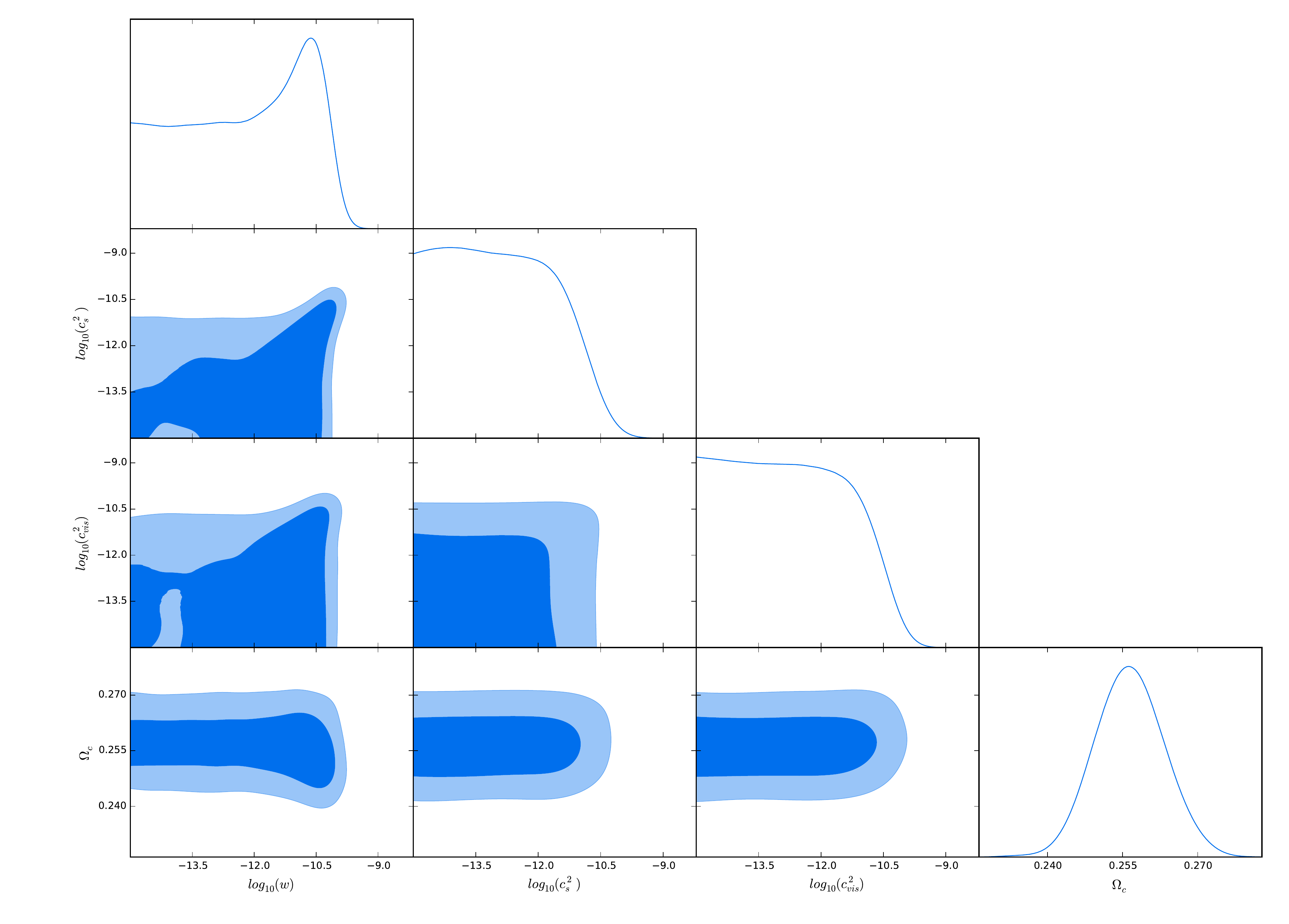}}}
\caption{The 2D $68\%$, $95\%$ confidence contours and the 1D marginalized posterior distributions for the parameters $(w_0,\csz, \cvisz, \Omega_c)$ of the initially relativistic model for the Planck (without trispectrum), WL, BAO and SNIa data.\label{fig:reltest3}}
\end{figure*}

\begin{figure*}[!t]
\centering \vspace{0cm}\rotatebox{0}{\vspace{0cm}\hspace{0cm}\resizebox{1\textwidth}{!}{\includegraphics{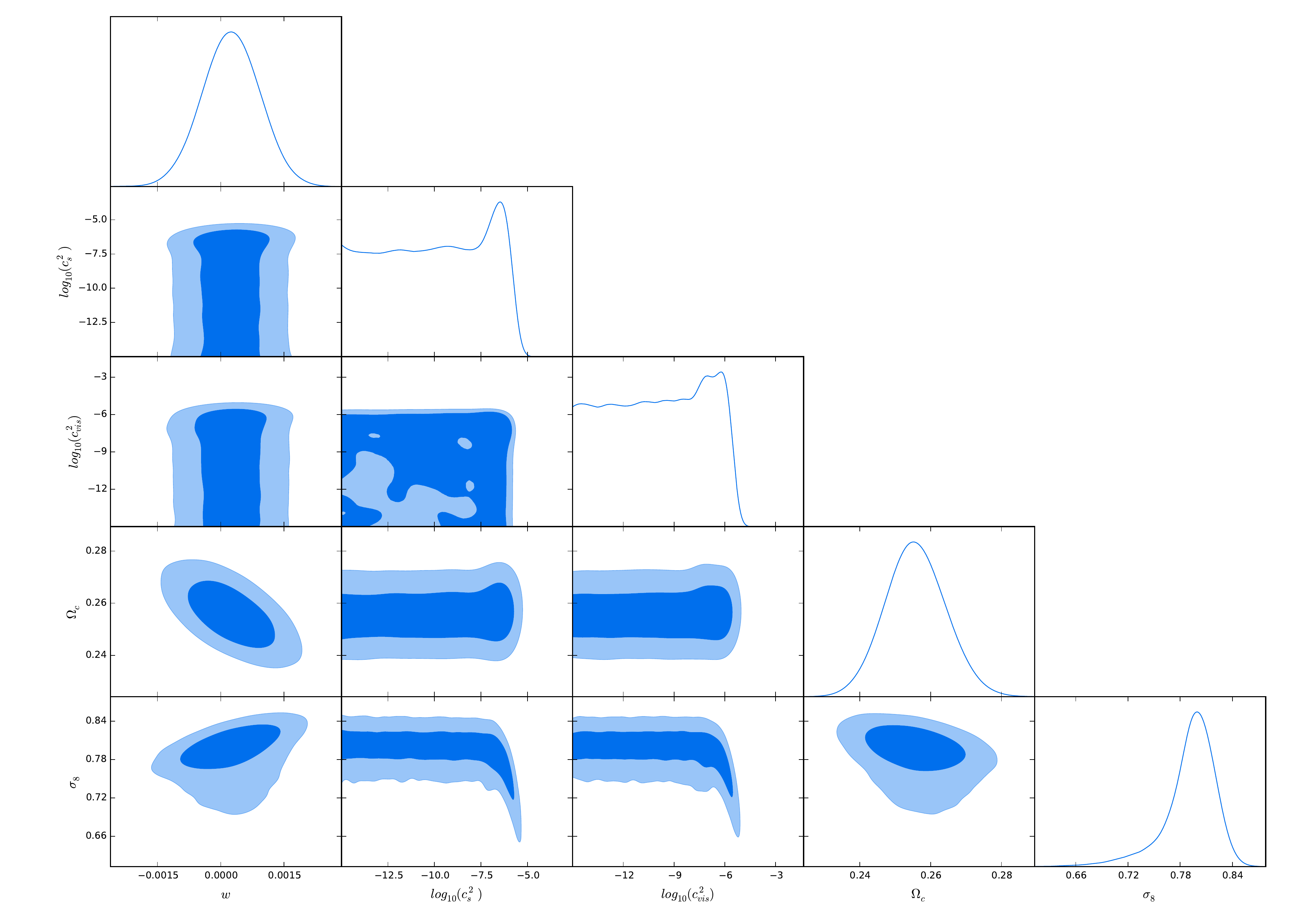}}}
\caption{The 2D $68\%$, $95\%$ confidence contours and the 1D marginalized posterior distributions for the parameters of the model  $(w,\cs, \cvis,\Omega_c)$ and the derived amplitude $\sigma_{8}$ in the constant parameterization for the Planck (without trispectrum), WL, BAO and SNIa data.	\label{fig:test3a}}
\end{figure*}

\section{Extra Plots}
In this section we show for completeness two extra plots, both for the case when we include the WL data to Planck and distance data by using the initially relativistic and constant parametrizations. In Fig.~\ref{fig:reltest3} we show the 2D $68\%$, $95\%$ confidence contours and the 1D marginalized posterior distributions for the parameters $(w_0,\csz, \cvisz, \Omega_c)$ of the initially relativistic model for the Planck (without trispectrum), WL, BAO and SNIa data, while in Fig.~\ref{fig:test3a} we show the 2D $68\%$, $95\%$ confidence contours and the 1D marginalized posterior distributions for the parameters of the model $(w,\cs, \cvis,\Omega_c, \sigma_{8,0})$ in the case when the first three of these are constant and free to vary, for the Planck (without trispectrum), WL, BAO and SNIa data. As mentioned in the main text, the constraints that result from the addition of the WL are practically the same as without it, as the fit is mainly driven by Planck and the distance probes.

\bibliographystyle{utcaps}
\bibliography{dmprop}

\end{document}